\documentclass[a4paper,twocolumn,noarxiv,accepted=2022-05-16,11pt]{quantumarticle}
\pdfoutput=1
\usepackage[utf8]{inputenc}
\usepackage[english]{babel}
\usepackage[T1]{fontenc}
\usepackage{amsmath,amsfonts,amsthm,amssymb,mathtools}
\usepackage{bbm,bm}
\usepackage{hyperref}
\usepackage[numbers,sort&compress]{natbib}
\usepackage{hyperref}
\usepackage{dsfont}

\newcommand{\ket}[1]{| #1 \rangle}
\newcommand{\bra}[1]{\langle #1 |}
\newcommand{\kb}[2]{| #1 \rangle\hspace{-2pt}\langle #2 |}
\newcommand{\Al}{\mathcal{A}}
\newcommand{\Bl}{\mathcal{B}}

\newtheorem{theorem}{Theorem}

\newtheorem{result}[theorem]{Result}

\renewcommand{\Pr}{\mathrm{P}}

\newcommand{\openone}{\mathbbm{1}}

\begin{document}
\title{Quantum Bell inequalities from Information Causality -- tight for Macroscopic Locality}
\author{Mariami Gachechiladze}
\thanks{mgachech@uni-koeln.de}
\affiliation{Institute for Theoretical Physics, University of Cologne, 50937 Cologne, Germany}

\author{Bart\l{}omiej B\k{a}k}
\affiliation{Department of Mathematical Methods in Physics, Faculty of Physics, University of Warsaw, ul.~Pasteura 5, 02-093 Warsaw, Poland}

\author{Marcin Paw{\l}owski}
\affiliation{International Centre for Theory of Quantum Technologies (ICTQT), University of Gdansk, 80-308 Gda\'nsk, Poland}

\author{Nikolai Miklin}
\thanks{nikolai.miklin@ug.edu.pl}
\affiliation{International Centre for Theory of Quantum Technologies (ICTQT), University of Gdansk, 80-308 Gda\'nsk, Poland}

\date{\today}
\begin{abstract}
In a Bell test, the set of observed probability distributions complying with the principle of local realism is fully characterized by Bell inequalities. Quantum theory allows for a violation of these inequalities, which is famously regarded as Bell nonlocality. However, finding the maximal degree of this violation is, in general, an undecidable problem. Consequently, no algorithm can be used to derive quantum analogs of Bell inequalities, which would characterize the set of probability distributions allowed by quantum theory.
Here we present a family of inequalities, which approximate the set of quantum correlations in Bell scenarios where the number of settings or outcomes can be arbitrary. We derive these inequalities from the principle of Information Causality, and thus, we do not assume the formalism of quantum mechanics. Moreover, we identify a subspace in the correlation space for which the derived inequalities give the necessary and sufficient conditions for the principle of Macroscopic Locality. As a result, we show that in this subspace, the principle of Information Causality is strictly stronger than the principle of Macroscopic Locality.
\end{abstract}
\maketitle

\section{Introduction}
At the most general level, experimenters' interaction with physical apparatuses is described by the probability distribution of their responses to the operations performed. To predict which types of distributions occur in nature is an important problem from both a conceptual and applied standpoints. A Bell test is the most central and relevant scenario from both of these perspectives~\cite{bell1964einstein}. A phenomenon known as nonlocality, that is demonstrated in a Bell test, implies that the principle of local realism~\cite{einstein1935can}, which appears to be inherent to all physical systems, must be rejected~\cite{bell1964einstein}. At the same time, nonlocality promises unconditional communication security~\cite{acin2007device}, randomness amplification~\cite{pironio2010random} and may even be the source of quantum computational advantage~\cite{bravyi2018quantum}.

To demonstrate nonlocality one needs to show that the observed correlations cannot be reproduced by local hidden variable models, which is the name for probabilistic theories satisfying the principle of local realism~\cite{bell1964einstein}. Theoretically, this task can be easily achieved since the set of distributions corresponding to local models is fully characterized by Bell inequalities, and quite a few examples of them are present in the literature~\cite{clauser1969proposed,collins2002bell,collins2004relevant,pironio2005lifting,rosset2014classifying,cruzeiro2019complete, gachechiladze2017completing}. Violation of a Bell inequality implies that the observed distribution is nonlocal, which has also been experimentally demonstrated~\cite{shalm2015strong,hensen2015loophole,giustina2015significant}. Although classification of all Bell inequalities remains a challenge, there is a systematic way in which new relevant Bell inequalities can be identified~\cite{pitowsky1991correlation}.

A much harder problem is characterizing the set of probability distributions attainable in quantum theory. No algorithm or systematic method is known to derive quantum analogs of Bell inequalities, especially non-linear ones, and consequently a very few examples of relevant quantum Bell inequalities are known~\cite{tsirelson1987quantum,landau1988empirical,uffink2002quadratic,zohren2010tight,yang2011quantum}. Even the most straightforward approach of translating ``classical'' Bell inequalities by substituting their local hidden variable bounds by the maximum attainable in quantum theory is only possible in very limited cases~\cite{cirelson1980quantum} due to the difficulty of the problem~\cite{ji2020mip}. Important results in this direction come from a fast-growing field of quantum self-testing~\cite{mayers2003self}. The goal of self-testing protocol is to uniquely (up to local isometries) identify a quantum state and (sometimes) corresponding quantum measurements, solely from the observed correlations in a Bell test. Often, the self-testing argument is formulated in the form of the maximal violation of a Bell inequality attainable in quantum mechanics. The latter value precisely singles out the self-tested state and measurements~\cite{coladangelo2017all}. On the other hand, such a quantum
bound on the Bell inequality leads to a linear quantum Bell inequality (see Re.~\cite{vsupic2020self} for a comprehensive review).

In this work, we present an infinite family of quantum Bell inequalities, that generalize and outperform some previously known results~\cite{cirelson1980quantum,uffink2002quadratic}. These inequalities are derived from Information Causality~\cite{pawlowski2009information}, a physical principle based on a set of information-theoretic assumptions that are satisfied by quantum theory but do not rely on its full formalism. We show that these inequalities are also implied by the principle of Macroscopic Locality~\cite{navascues2010glance}, and, moreover, we identify a subspace in the set of probability distributions in which the derived inequalities also constitute the necessary condition for this principle.

Bounding the set of quantum correlations from physical principles offers several advantages. First, a physical principle might lead to a much more efficient description of the set of correlations compatible with it, at the expense of giving less tight approximation of the quantum set. We give a few examples below in which this trade-off is apparent. Second, since the axioms of quantum theory are not physically motivated, the problem of identifying appropriate principles that lead to the same set of correlations as quantum theory is of foremost importance. Information Causality is possibly the most intriguing of the known principles, since it has not been established yet whether it can be satisfied by any correlations stronger than quantum. Finally, if quantum theory is eventually replaced with a more general probabilistic theory~\cite{rovelli2004quantum}, those inequalities that are based on physical principles may still be valid.

Few examples of quantum Bell inequalities resulting from different physical principles already appear in the literature. One of the most fundamental principles,  the principle of no-signaling~\cite{popescu1994quantum}, allows for an efficient description of the set of correlations compatible with it. However, this approximation tuns out to be too coarse since it allows for correlations saturating the algebraic maximum of $4$ of Clauser-Horne-Shimony-Holt (CHSH) inequality~\cite{clauser1969proposed}, while the maximal value permitted by quantum theory is $2\sqrt{2}$~\cite{cirelson1980quantum}. The principle of Information Causality (IC)~\cite{pawlowski2009information}, on the other hand, recovers the so-called Tsirelson bound of $2\sqrt{2}$. However, the set of correlations compatible with this principle is very hard to describe, due to its formulation. The IC principle constraints the amount of mutual information between a remote data and its guess in a communication game~\cite{pawlowski2010entanglement} (See Section~\ref{ICPrinciplesSection} for details).
Despite this indirect formulation, surprisingly, a simple quadratic quantum Bell inequality, known as Uffink inequality~\cite{uffink2002quadratic} was shown to follow from the IC principle~\cite{pawlowski2009information}. The principle of Macroscopic Locality (ML)~\cite{navascues2010glance} offers a direct and relatively simple formulation in terms of a semidefinite program, which also leads to the correct bound on CHSH inequality. However, this formulation has a major drawback that it requires solving an optimization problem. Since the number of optimization variables grows with the size of the probability vector, this means that an analytical description of the set of macroscopic local correlations e.g., in terms of inequalities, is possible only in simple scenarios~\cite{navascues2007bounding,yang2011quantum}.

In this paper, we generalize the aforementioned results by presenting an infinite family of quantum Bell inequalities in which the number of measurement settings or outcomes can be taken arbitrary high. We show that these inequalities, which take a simple form of quadratic expressions in terms of the observed probabilities, follow from both the IC and ML principles. This is especially intriguing since the two principles have very district formulations. The interplay between the IC and ML principle was investigated earlier by giving examples of correlations that satisfy the ML principle but violate the principle of IC~\cite{cavalcanti2010macroscopically}. In this work we identify an entire subspace of the correlation space in which the derived inequalities give the exact boundary of the set of macroscopically local correlations, or in other words provide the necessary and sufficient condition for the ML principle. It means that everywhere in this subspace, which is defined by the symmetries of the derived inequalities, the principle of IC is at least as strong as the principle of ML.

Finally, in this work, we develop systematic methods that can be used to derive quantum Bell inequalities from the both considered physical principles. Concerning derivations from the IC principle, the main challenge is in calculating the guessing probabilities resulting from using multiple copies of the same probability distribution, the process called concatenation~\cite{pawlowski2009information}. We show that for a given communication protocol, this process can be formulated as a Markov chain. Using a tool known as z-Transform~\cite{howard1960dynamic,oppenheim2001discrete} we derive the solution of a special type of Markov chains (originating from the communication protocol) of growing dimension.
We believe that the derivations related to the Markov chains are elegant and can be of independent interest.  For obtaining the characterization of the set corresponding to the ML principle, the main problem that one faces is the growing number of optimization variables. In this work we show that if one considers a subspace of the correlation space defined by some symmetry group, the condition of ML can be broken into a set of conditions of lower dimensions, each of which can be solved analytically. We provide an explicit construction for a specific symmetry group in this paper, but the overall idea should be easily generalizable to other groups. 

The main part of the paper is divided into two parts. In the first part, we give the formal statement of the IC principle and present our quantum Bell inequalities with a high-level proof. In this proof, an explicit solution of a particular Markov chain is given. In the second part, we give a formulation of the principle of ML as a semidefinite program and show that the derived inequalities constitute necessary and sufficient conditions for the considered subspace of correlations. Technical details are left to Appendix.

\begin{figure}[t!] \centering
		\includegraphics[width=.38\textwidth]{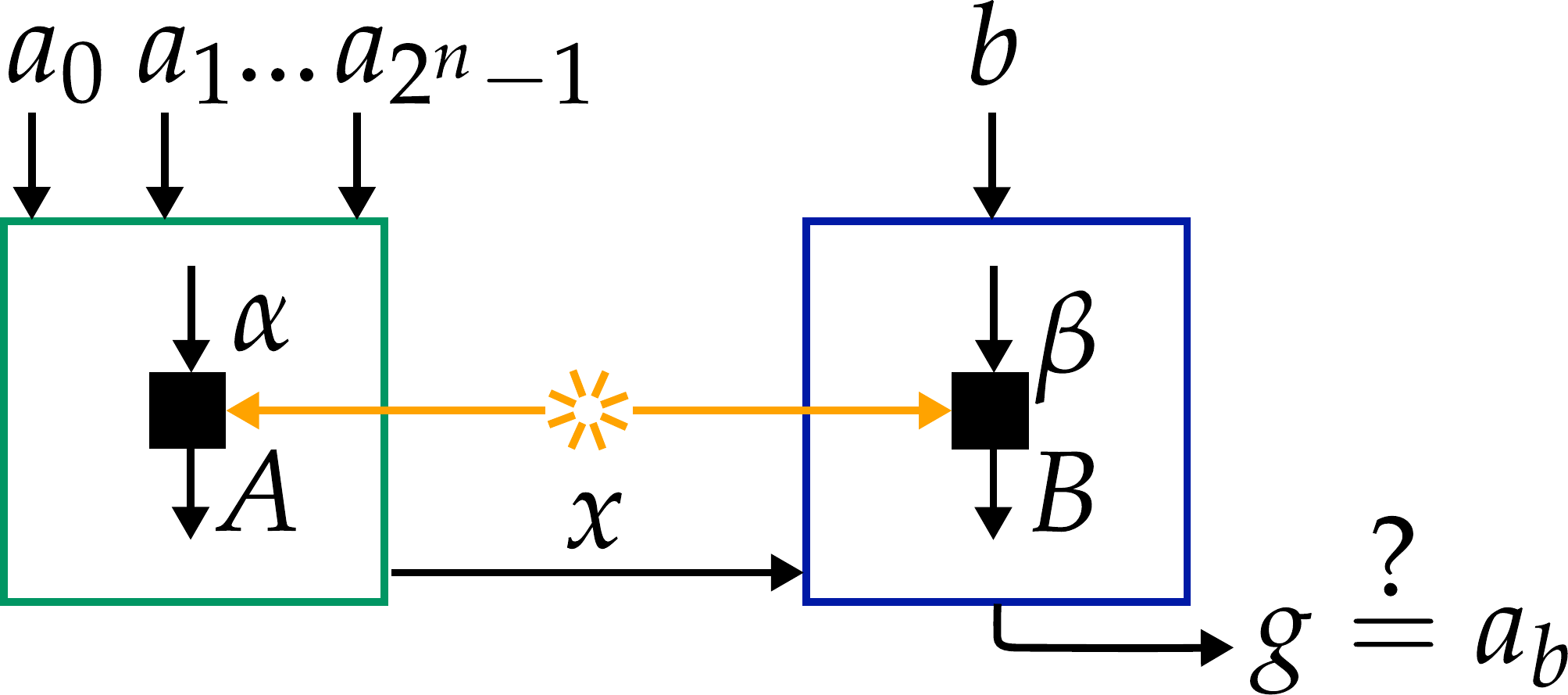}
		\caption{Information Causality scenario. Alice receives a data set of random dits $\{a_0,a_1,\dots,a_{2^n-1}\}$, while Bob is given $b$, which takes values between $0$ and $2^n-1$. Alice and Bob have access to devices which produce nonsignaling correlations with inputs denoted by $\alpha$ and $\beta$ and outputs by $A$ and $B$. Bob's guess of $a_b$ is denoted by $g$. To facilitate with the guessing, Alice sends a classical message $x$ usually taken to be of the same size as the data variables $a_i$.}
		\label{fig:scenario}
\end{figure}

\section{Notations}
Here we list some of the notations used throughout the paper. We use upper-case and lower-case letters as well as Greek letters to denote random variables. To avoid possible confusions, we always write explicitly when we introduce a random variable. For the same reason, for events such as a random variable $A$ taking a value $k$, i.e., $A=k$, or two random variables $A$ and $B$ taking equal values, i.e., $A=B$, the probability is always spelled out as $\Pr(A=i)$ or $\Pr(A=B)$. All other notations resembling probabilities, e.g., $p(e|j)$ where  $e$ and $j$ are indices (not random variables), should be treated formally as functions.
For correlations observed in a bipartite Bell scenario, with inputs $\alpha$ and $\beta$ and outputs $A$ and $B$, the observed correlations are denoted as $\Pr(A=k,B=l|\alpha=i,\beta=j)$ in accordance with our general notation of probabilities.

The addition modulo $d$ is denoted by $\oplus$ and we use $\overline{x}$ to denote the inverse (or inverse random variable) of $x$, i.e., $\overline{x} {\oplus}x = 0$. We also use $x\ominus y$ to denote $x\oplus\overline{y}$. We use the convenient Dirac notation $\ket{i}$, $i\in\{0,1,\dots d\}$ for the basis vectors in $\mathds{R}^d$. We also use $\ket{\psi}$ for a general, not necessary normalized column vector $\psi$ in the same space $\mathds{R}^d$. We use the common notations $[n]\equiv \{0,1,\dots,n-1\}$, $\mathrm{i} = \sqrt{-1}$, $x^*$ for the complex conjugate of $x\in \mathds{C}$, $\Re(x)$ and $\Im(x)$ for the real and imaginary parts of $x$, and $\omega=\exp{(2\pi \mathrm{i}/d)}$ for the $d$-th root of unity. Finally, we abbreviate a $d$-dimensional digit as ``dit''.

\section{Information causality principle}\label{ICPrinciplesSection}
We start by summarizing the statement of Information Causality (IC) principle~\cite{pawlowski2009information}. The IC is defined in communication scenario, known as an entanglement-assisted random access code~\cite{pawlowski2010entanglement}, see Fig.~\ref{fig:scenario}. Two parties, Alice and Bob, face a communication challenge.  Alice is given a data set of $2^n$ $d$-outcome random variables, which we denote as $\{a_0,a_1,\dots,a_{2^n-1}\}$ ($a_i$ takes values in $[d]$ for all $i\in [2^n]$, $n$ is a positive integer). Bob is asked to guess one of the dits from the data set given to Alice, with the choice determined by another random variable $b$ which takes values in $[2^n]$. All the random variables $a_0,a_1,\dots,a_{2^n-1}$, and $b$ are considered independent and uniformly distributed. Their values are drawn at random before each run of the experiment. In each run, Alice communicates a message $x$ to Bob with possible information about some of her data. At the end of each run, Bob outputs a value of his guess $g$, which also takes values in $[d]$. Both parties succeed if $g=a_b$, by which we mean (with a slight abuse of notation) that $g=a_y$, while $b=y$, $\forall y\in [2^n]$. 

Clearly, when the size of the message $x$ is smaller than the size of the data set, $\{a_0,a_1,\dots,a_{2^n-1}\}$ no perfect guessing is possible. This intuition, however, fails if the parties are allowed to facilitate their communication with extreme nonsignaling correlations, also known as PR-boxes~\cite{popescu1994quantum}. In fact, Ref.~\cite{pawlowski2009information} (and earlier works by van Dam~\cite{van2005implausible,vandam2013implausible})  demonstrated that, if parties have access to $(2^n-1)$ PR-boxes, a single bit of communication (for the case of $d=2$) is enough to win the game with the unit probability. The main idea there is that Alice uses the output(s) $A$ of the PR-box(es) for the message encoding. By using output(s) $B$, Bob can effectively choose which bit of Alice to decode. 

The IC principle aims at restricting such non-physical behaviors based on the set of natural assumptions also satisfied by quantum theory. More precisely, it states that for nonsignaling theories equipped with a measure of correlations, $I(\cdot;\cdot)$, that satisfies the chain rule, data processing inequality, and which reduces to the Shannon mutual information for classical variables, the following inequality holds~\cite{pawlowski2009information},
\begin{equation}
    \sum_{y=0}^{2^n-1}I(a_y;g|b=y)\leq |x|,
    \label{eq:ic_statement}
\end{equation}
where $|x|\equiv I(x;x)$ is the size of the message that Alice sends to Bob. Equivalently, the bound on the right-hand side of Eq.~(\ref{eq:ic_statement}) can be seen as the capacity of the identity channel used for communication~\cite{pawlowski2016information}. The statement in Eq.~(\ref{eq:ic_statement}) can be understood in the following way. Even if utilizing nonsignaling correlations, in the discussed communication game, provides an advantage in terms of guessing probability, there should be no benefit in terms of information gain.

Considering the formulation of Eq.~(\ref{eq:ic_statement}), it is not clear how to use it to bound the set of correlations in a Bell test. The reason for this is that one needs to consider all possible ways Alice and Bob can utilize the correlations produced by, generally, an infinite number of copies of their devices in the IC game. As a result, no characterization of the set of correlations complying with the IC principle is known. Moreover, it is still possible that the IC principle can single out the set of quantum correlations, and the description of the latter is known to be an undecidable problem~\cite{ji2020mip}. Despite these difficulties, in the original work of Ref.~\cite{pawlowski2009information} it was shown that in the case of binary inputs and outputs, i.e., for $\Pr(A=k,B=l|\alpha=i,\beta=j)$ with $k,l,i,j\in [2]$, a simple quantum Bell inequality, know as Uffink inequality~\cite{uffink2002quadratic} can be derived from the IC principle. This approximation in terms of Uffink inequality is tight in the sense that it predicts the Tsirelson's bound of $2\sqrt{2}$ as the maximum value of CHSH expression attained in quantum mechanics~\cite{tsirelson1987quantum,clauser1969proposed}.

In this work, we consider a more general Bell scenario, in which the number of settings or outcomes can be arbitrary high. More precisely we look at the distributions $\Pr(A=k,B=l|\alpha=i,\beta=j)$, with $k,l,i\in [d]$, and $j\in[2]$, for any possible $d$.
In order to state our main result, let us define the following linear expressions of the observed correlations $\Pr(A=k,B=l|\alpha=i,\beta=j)$, with $k,l,i\in [d]$, and $j\in[2]$,
\begin{equation}
    \label{eq:pej}
    p(e|j)=\frac{1}{d}\sum_{i=0}^{d-1}\Pr(A \oplus B = i\cdot j\oplus e|\alpha=i,\beta=j),
\end{equation}
for all $e\in [d]$ and $j\in [2]$. We use the lower-case letter $p$, which shall not be confused for the probability $\Pr$ (see Notations). These functions play the role of guessing probabilities for each individual pair of the devices, however, for the moment they should be understood simply as a notation convenient for presenting the following result.

\begin{result}\label{th:ic}
Any nonsignaling theory satisfying the Information Causality principle complies with the following inequalities in $\{p(e|j)\}_{e,j}$ for all $i \in \left\{1,\dots,\left\lfloor{d}/{2}\right\rfloor\right\}$ and $\omega=e^{{2 \pi \mathrm{i} }/{d}}$,
\begin{equation}
\label{eq:main}
    \left|\;\sum_{e=0}^{d-1}p(e|0)\omega^{e \cdot i}\;\right|^2+ \left|\;\sum_{e=0}^{d-1}p(e|1)\omega^{e \cdot i}\;\right|^2\leq 1.
\end{equation}
\end{result}
\noindent Notably, for $d=2$, the inequality in Eq.~(\ref{eq:main}) reduces to the famous quadratic quantum Bell inequality by Uffink~\cite{uffink2002quadratic}.

\begin{proof}
The proof consists of several parts: \textit{(i)~Communication protocol} -- Here a detailed description of IC scenario and concatenation protocol are given;  \textit{(ii)~Guessing probability} -- In this part, we express the success and error probabilities of Bob's guesses in terms of the nonsignaling correlations shared between the parties; \textit{(iii)~New inequalities from IC} -- In this final part of the proof we connect the IC statement with the guessing probabilities from part \textit{(ii)}, which leads us to our main result.

\textit{(i) Communication protocol.} -- Alice and Bob are in the IC scenario (see Fig.~\ref{fig:scenario}). They have access to a pair of devices producing nonlocal correlations, or as we call it, nonsignaling box (NS-box). We denote the inputs of the shared NS-box by $\alpha$ and $\beta$ and outputs by $A$ and $B$, respectively (see Notations). In this work we focus on $d2dd$ NS-boxes, i.e., $\beta$ is binary and $\alpha,A$, and $B$ take values in $[d]$.  Then, as the communication protocol for Alice and Bob, we can choose the direct generalization of the case $2222$~\cite{van2005implausible}, where Alice communicates to Bob a single dit message $x$, and
\begin{align}\label{eq:protocol_1}
\begin{split}
    \alpha & = \overline{a}_0 {\oplus} a_1,\; x = a_0 {\oplus} A,\\
    \beta & = b,\; g = x {\oplus} B.
\end{split}
\end{align}
If Alice and Bob share an NS-box, which exhibits the correlations $A\oplus B = \alpha\cdot\beta$, then Bob's guesses are perfect, i.e., $g=a_0$, if $b=0$ and $g=a_1$ if $b=1$. Consequently, such NS-boxes violate the IC principle, as the left-hand side of Eq.~(\ref{eq:ic_statement}) is $2$ while $|x|=1$. More generally, Eq.~(\ref{eq:ic_statement}) puts an upper bound on the probability $\Pr(A\oplus B = \alpha\cdot\beta)$. However, for the above protocol for $n=1$, i.e., two dits of Alice's data, such bounds turn out to be loose, and they become tighter when $n$ increases~\cite{pawlowski2009information}. For example, in $d=2$ case, the bound one obtains from IC for $n=1$ is approximately $0.8889$, while the true quantum value is $1/2+1/2\sqrt{2}\simeq 0.8536$~\cite{cirelson1980quantum}, which one gets in the limit of $n\rightarrow\infty$~\cite{pawlowski2009information}.

The protocol in Eq.~(\ref{eq:protocol_1}) can be generalized to an arbitrary $n$ through the process called \textit{concatenation}~\cite{pawlowski2009information}. In the concatenation protocol, outputs of some boxes are fed as inputs of subsequent ones, forming a layered structure in this way. We  use the superscript indices for inputs and outputs of a box to specify the layer to which this NS-box belongs to. There are $n$ layers of NS-boxes in total with $2^{m}$ boxes on each $m$-th layer, making it $2^n-1$ boxes for the whole protocol. For convenience, we enumerate the layers, $m$ in descending order from $(n-1)$ to $0$. 

We start with generalizing Alice's encoding strategy. First, we group the data set $\{a_0,a_1,\dots,a_{2^n-1}\}$ in consecutive pairs, i.e., $(a_0,a_1),(a_2,a_3),\dots,(a_{2^n-2},a_{2^n-1})$. We then assign to each pair an NS-box with Alice's output $A^{n-1}_i$. Here, the superscript $(n-1)$ stands for the layer and the subscript $i\in[2^{n-1}]$ distinguishes NS-boxes within that layer. The input $\alpha^{n-1}_i$ to the $i$-th box is $(\overline{a}_{2i}\oplus a_{2i+1})$, which resembles the protocol in Eq.~(\ref{eq:protocol_1}). At the end of the first step of encoding, we form $2^{n-1}$ new dits $a_{2i}\oplus A_{i}^{n-1}, i\in[2^{n-1}]$, which serve as the data set for the $(n-2)$-th layer of NS-boxes. In other words, what used to be the message $x$ in Eq.~(\ref{eq:protocol_1}) plays the role of the data set on the $(n-2)$-th layer of concatenation. The encoding strategy then proceeds iteratively, the size of the data set halving on each lower layer of concatenation, until there are only two dits of data on the $0$-th layer. The output of the $0$-th layer of the encoding strategy is the message $x$ that Alice sends to Bob.

The decoding strategy of Bob is best to be understood in terms of walks. First, let us introduce $\vec{b}=(b_0,b_1,\dots,b_{n-1})$, the binary form of $b$, i.e., $b=\sum^{n-1}_{m=0} b_m 2^m$. Each bit $b_m$ dictates the ``left" $(0)$ or ``right" $(1)$ direction of the $m$-th step that Bob needs to take in unscrambling the $m$-th layer of Alice's encoding. In the $0$-th step of his decoding, Bob inputs $\beta^0_0 = b_0$ to the only layer-$0$ NS box and adds his output dit $B^0_0$ to the message $x$. If now $b_1=0$, he ``moves left" (by which we mean that he uses $0$-th box of layer $1$ with an input $\beta^1_0=0$) and he adds $B^1_0$ to $x\oplus B^0_0$. Alternatively, if $b_1=1$, Bob ``moves right" (i.e., he inputs $\beta^1_1 = 1$ to the $1$-st box of layer $1$) and he adds $B^1_1$ to $x\oplus B^0_0$. He proceeds by moving up the layers till the $(n-1)$-th layer,  using only one box $i_m$ per each layer $m$. This way, on the $m$-th layer with totally $2^m$ boxes, Bob chooses $i_m$-th box, where $i_m=2^{m}b_m+i_{m-1}$ can be calculated recursively for $1\leq m\leq (n-1)$. Finally, the sum of the outputs, $g=x\oplus B^0_0\oplus\dots$ is the guess $g$ that Bob makes.

For reader's convenience, we give an explicit form of encoding-decoding strategy for $n=2$ below:
\begin{align}\label{eq:protocol_2}
\begin{split}
    \alpha^1_0 & = \overline{a}_0{\oplus} a_1,\;\alpha^1_1 = \overline{a}_2{\oplus} a_3,\\
     \alpha^0_0 & = \overline{(a_0{\oplus} A^1_0)}\oplus (a_2{\oplus} A^1_1), \\
    x & = (a_0{\oplus} A^1_0)\oplus A^0_0,\\
    \beta^0_0 & = b_0,\; \beta^1_0 = b_1,\; \beta^1_1 = b_1,\\
     g & = x{\oplus} B^0_0\oplus B^1_{b_0},
    \end{split}
\end{align}
where by $B^1_{b_0}$ we mean $B^1_0$, if $b_0=0$, and $B^1_1$, otherwise.

\textit{(ii) Guessing probability.} --
Here we work out the guessing probabilities $\{\Pr(g=a_y|b=y)\}_y$ from the concatenation protocol described in (i), and also the probabilities of errors. We define the  probability of making an error $E$ for the fixed value of $b$,
\begin{equation}\label{eq:final_prob}
    P(E|y)\equiv \Pr(g=a_y\oplus E|b=y),
\end{equation}
$\forall E\in [d],\forall y\in [2^n]$ where $P(0|y)$ is the probability of Bob guessing $a_y$ correctly. For $d=2$ there is only one probability of error, which does not need to be calculated as $P(1|y)=1-P(0|y)$.  However, for the further derivations, for $d>2$, we need to calculate  error probabilities in Eq.~(\ref{eq:final_prob}) for all $E$. 

In our derivations, we assume that all $(2^n-1)$ NS-boxes are identical, i.e., they all exhibit the same correlations. For the simplest case, $n=1$, by employing the protocol in Eq.~(\ref{eq:protocol_1}), we can explicitly calculate the probability $P(E|y)$,
\begin{align}
    P(E|y)& =
    \Pr(x\oplus B=a_y\oplus E|b=y) \nonumber\\
    &=\Pr(a_0\oplus A\oplus B=a_y\oplus E|b=y) \nonumber\\ 
    &=\Pr(A\oplus B=a_y\oplus \overline{a}_1\oplus\alpha\oplus E|b=y) \nonumber \\
    & = p(E|y).
\end{align}
In the last equality, on the right-hand side we have the correlation of the NS-box defined in Eq.~(\ref{eq:pej}) and the equality holds since $a_y\oplus \overline{a}_1\oplus\alpha = \alpha\cdot y$ for all $y~\in~[2]$.

Our goal is to compute $P(E|y)$ for an arbitrary $n$, however, this proves to be an intricate task. The largest effort is to count the resulting error $E$ as a sum of the individual errors $e^m$ corresponding to the  NS-box used on every $m$-th layer. To give an example,  if we use the protocol in Eq.~(\ref{eq:protocol_2}) for $n=2$, we get two errors: $e^0$ and $e^1$ that sum up to $E=e^0\oplus e^1$. Furthermore, there are $d$ possible pairs of $e^0$ and $e^1$ that sum to the same resulting error $E$.  More generally, for $n$ boxes, we have $d^{n-1}$ combinations of errors, and we have to account for all of them in the final guessing probability. 

Finally, apart from listing all combinations of errors $\{e^m\}_{m=0}^{n-1}$ which contribute to $E$, we are also interested in the type of boxes used, i.e., the value of $\beta$ inserted in the box that produces $e^m$. For such an error-handling, it is again beneficial to think of Bob's decoding strategy as a walk. 

By looking at his input $b=y$, Bob finds that the guess of the value of $a_y$ involves aiming $(n-k)$ times at the left (inserting $\beta=0$) and $k$ times at the right (inserting $\beta=1$), where $k$ is the weight of $y$, i.e., the number of $1$s in the binary form of $y$.
Since the addition ${\oplus}$ is symmetric and associative, every $P(E|y)$ with input of the same weight is going to be the same. Hence, in full generality, we can redefine $P(E|y)$ in Eq.~(\ref{eq:final_prob}) as $P(E|k)$, with $k\in [n]$. 

In order to calculate $P(E|k)$, we introduce one more notation
\begin{equation}
\ket{Q^{(k)}(j)}=\sum_{e=0}^{d-1}Q^{(k)}_e(j)\ket{e},\quad j\in [2],
\end{equation}
where $Q^{(k)}_e(j)$ is the probability of making an error $e$ when aiming $k$ times in $j$-th direction, and $\{\ket{e}\}_{e=0}^{d-1}$ is the computational basis in $\mathds{R}^d$. If we also consider the errors accumulating when aiming $(n-k)$ times in $\bar{j}$ (opposite to $j$) direction, we get the total probability for each error $E$,
\begin{equation}\label{eq:final_prob_k}
    P(E|k)= \bra{Q^{(k)}(0)}S_{E}\ket{Q^{(n-k)}(1)},
\end{equation}
where $S_{E}=\sum_{i=0}^{d-1}\kb{i}{\overline{i}\oplus E}$ is a permutation matrix. 

Let us consider some examples of $Q_e^{(k)}(j)$ for $k=1$ and $k=2$ and arbitrary $d$. Starting from the simplest case $k=1$, the probability of making $e$ errors, when aiming one time in $j$-th direction is $Q_e^{(1)}(j)=p(e|j)$, $\forall e\in[d]$. For $k=2$, we have
\begin{equation}
    Q_e^{(2)}(j)=\sum_{{e'}=0}^{d-1}p(e\oplus e'|j)p(\bar{e'}|j).
\end{equation}
More generally, the error probability vector, $\ket{Q^{(k)}(j)}$ can be expressed as the $k$-th state vector in the Markov process with the transition matrix depending on $j$,
\begin{equation}\label{eq:tran_matr}
    M_j = \sum_{e=0}^{d-1}\sum_{e'=0}^{d-1}p(e'\oplus \bar{e}|j)\kb{e'}{e}.
\end{equation}
In other words, we can obtain the vector for $k=2$ from $k=1$ as $\ket{Q^{(2)}(j)} =M_j\ket{Q^{(1)}(j)}$, and more generally, 
\begin{equation}
    \ket{Q^{(k)}(j)} = M^k_j\ket{0}.
\end{equation}
Hence, we have reduced the problem to finding $M^k_j$, the $k$-th power of the doubly stochastic matrix $M_j$, which can be managed using the formalism of the \textit{z-Transforms}~\cite{howard1960dynamic,oppenheim2001discrete}. With the help of z-Transform formalism, we calculated arbitrary $k$-th power of $M_j$ for small dimensions. Luckily, already from working out the cases of $d=3$ and $d=4$  we were able to conjecture the general form of $\ket{Q^{({k})}(j)}$ for any $d$, which we give below.

For $d$ odd, $e\in [d]$, the entries of $\ket{Q^{({k})}(j)}$, amount to
\begin{equation}
    \label{eq:Q_odd}
Q^{(k)}_{e}(j) = \frac{1}{d}+\frac{2}{d}\sum_{i=1}^{\frac{d-1}{2}}\Re{\left(l^k_i(j)\omega^{e\cdot i}\right)},
\end{equation}
where $l_i(j) = \sum_{e=0}^{d-1}p(e|j)\omega^{-e\cdot i}$ is a complex-valued function of the probabilities $p(e|j)$. 
The formula has a concise proof by induction (See Appendix~\ref{app:Q_induction}). A slight variation of the above formula for $d$ even is also given in Appendix~\ref{app:Q_induction}.

Knowing $\ket{Q^{(k)}(j)}$, we can calculate the probabilities $P(E|k)$ in Eq.~(\ref{eq:final_prob_k}), 
\begin{align}
        P(&E|k)-\frac{1}{d}=\sum_{e=0}^{d-1}Q^k_{e}(0)Q^{n-k}_{\bar e\oplus E}(1)-\frac{1}{d}\\
        =& \frac{4}{d^2} \sum_{i_1,i_2=1}^{\frac{d-1}{2}}\sum_{e=0}^{d-1}\Re{\left(\omega^{i_1 e}l_{i_1}^k(0)\right)}\Re{\left(\omega^{i_2 (\bar e\oplus E)}l_{i_2}^{n-k}(1)\right)}\label{eq:probderivations2}\\
=&\frac{2}{d}\sum_{i=1}^{\frac{d-1}{2}}\left( \cos\left(\frac{2\pi E i}{d}\right)A_i-\sin\left(\frac{2\pi E i}{d}\right)B_i\right)\label{eq:probderivations3}.
\end{align}
Eq.~(\ref{eq:probderivations2}) looks cumbersome, but it simplifies greatly once we use Euler's formula for the root of unity and work with real and imaginary parts of $l_{i_1}^{k}(0)$ and $l_{i_2}^{n-k}(1)$. An important observation is that if one first takes the sum over $e$ in Eq.~(\ref{eq:probderivations2}), the sums of cosine and sine functions are nonzero if and only if $i_1=i_2$, which leads to Eq.~(\ref{eq:probderivations3}) (See Appendix~\ref{app:proofOfMain} for more detailed calculation). In Eq.~(\ref{eq:probderivations3}) we introduced the following useful notations, $A_i=\Re{\left(l_{i}^k(0)\right)}\Re{\left(l_{i}^{n-k}(1)\right)}-\Im{\left(l_{i}^k(0)\right)}\Im{\left(l_{i}^{n-k}(1)\right)}$
    and    
    $B_i=\Re{\left(l_{i}^k(0)\right)}\Im{\left(l_{i}^{n-k}(1)\right)}+\Im{\left(l_{i}^k(0)\right)}\Re{\left(l_{i}^{n-k}(1)\right)}.$
 Again, we do not give here the formula for $P(E|k)$ for $d$ even due to its similarity with the odd case. For the formula and its proof see Appendix~\ref{app:proofOfMain}.

\textit{(iii) New inequalities from IC. --}
In this work, we study the consequences of the IC principle for nonsignaling boxes (NS-boxes) with $d\geq 2$ outcomes. To this end, we assume the statement of IC as in Eq.~(\ref{eq:ic_statement}), where $I(a;b) = H(a,b)-H(a)-H(b)$ for two random variables $a$ and $b$, $|x| = H(x)$, and $H(\cdot)$ is the Shannon entropy, which for convenience we define with natural logarithm, i.e., $H(x) = \sum_{i}\Pr(x=i)\ln \Pr(x=i)$. 

Having computed general $\ket{Q^{({k})}(j)}$, the only missing part in our derivation is to connect the guessing probabilities with the IC statement. 
Ref.~\cite{pawlowski2009information} used Fano's inequality~\cite{fano1968transmission} to this end. Here, we use a similar inequality for two $d$-outcome random variables $a$ and $b$, where $a$ is uniformly distributed, $b = a\oplus e$,  and  $H(e) = -\sum_{i=0}^{d-1}\Pr(e=i)\ln \Pr(e=i)$,
\begin{equation}\label{eq:fano}
    I(a;b)\geq \ln d-H(e).
\end{equation}
See a short proof in Appendix~\ref{app:fano}.
If we take $|x| = \ln d$,  $H(E|b=y) = -\sum_{E=0}^{d-1}P(E|y)\ln P(E|y)$ and use the new inequality, we can approximate the terms, which appear in the formulation of IC in Eq.~(\ref{eq:ic_statement}).
\begin{equation}\label{eq:boundsonH}
    \sum_{y=0}^{2^n-1}H(E|b=y)\geq (2^n-1)\ln d.
\end{equation}
On the other hand, we can also upper bound the entropy with a polynomial function. In particular, the following inequality holds
\begin{equation}\label{eq:approx_entr}
    H(E|b=y) \leq \ln d - \kappa \sum_{E=0}^{d-1}\left( P(E|y)-\frac{1}{d}\right)^2
\end{equation}
for $\kappa = \frac{d^2(\ln d-1)+d}{(d-1)^2}>0$. See Appendix~\ref{app:approx_entr} for the proof.
In the part (ii), we redefined $P(E|y)$ as $P(E|k)$ (see Eq.~(\ref{eq:final_prob_k})), where $k$ is the weight of $y$. Then, using Eqs.~(\ref{eq:boundsonH},\ref{eq:approx_entr}) and the fact that $\binom{n}{k}$  terms corresponding to $b=y$ have the same weight, we arrive at the following approximation of the statement of the IC principle:
\begin{equation}\label{eq:ic_P(E|k)_2}
    \sum_{k=0}^{n}\binom{n}{k}\sum_{E=0}^{d-1}\left( P(E|k)-\frac{1}{d}\right)^2 \leq \frac{(d-1)^2\ln d }{d^2(\ln d-1)+d}.
\end{equation}
In our proof, we are not concerned with the exact bound on the right-hand side of  Eq.~(\ref{eq:ic_P(E|k)_2}), but rather with the fact that it does not grow with $n$. Interestingly, the right-hand side of Eq.~(\ref{eq:ic_P(E|k)_2}) is maximal for $d=12$ and approximately equals $1.331$, while the limit of $d\rightarrow \infty$ is equal to $1$.

We can now insert probabilities from Eq.~(\ref{eq:probderivations3}) into Eq.~(\ref{eq:ic_P(E|k)_2}). Using trigonometric identities for the sums over $E$ we obtain (See details in Appendix~\ref{App:theorem1prooffinalStep}),
\begin{align}\label{eq:theorem1prooffinalStep}
    \sum_{E=0}^{d-1}\left( P(E|k)-\frac{1}{d}\right)^2=& \frac{2}{d}\sum_{i=1}^{\frac{d-1}{2}}\left(A_i^2+B_i^2\right)\\
    =&\frac{2}{d}\sum_{i=1}^{\frac{d-1}{2}}\left|l_i(0)\right|^{2k}\left|l_i(1)\right|^{2(n-k)}.\nonumber
\end{align}
The binomial sum in Eq.~(\ref{eq:ic_P(E|k)_2}) then amounts to 
\begin{equation}
    \sum_{k=0}^{n}\scalebox{0.9}{$\displaystyle\binom{n}{k}$}\left|l_i(0)\right|^{2k}\left|l_i(1)\right|^{2(n-k)}=\left(\left|l_i(0)\right|^2+\left|l_i(1)\right|^2\right)^n
\end{equation}
Since this is true for any $i\in \{1,2,\dots,\left\lfloor{d}/{2}\right\rfloor\}$ and arbitrarily large $n$, our main results, the statement of Eq.~(\ref{eq:main}) follows. For even $d$ the derivations are very similar, and we include them in Appendix~\ref{App:theorem1prooffinalStep}.  
\end{proof}

In the next section, we first describe the set of correlations compatible with the Macroscopic Locality principle, and then compare the two results.

\section{Macroscopic Locality principle}\label{sec:ML}
The principle of Macroscopic Locality (ML) postulates that correlations resulting from measurements on microscopic systems should obey the laws of classical mechanics, in particular local realism~\cite{bell1964einstein}, if a large number of copies of these systems are considered, making them macroscopic~\cite{navascues2010glance}. We do not give the formal statement of the ML principle in this paper. Instead, for its quantitative formulation we use the fact that the set of macroscopic local correlations coincides with the first level, often denoted as Q$^1$, of the hierarchy of Ref.~\cite{navascues2007bounding}. Note that there exit other notions of ``macroscopicity'' in quantum mechanics~\cite{leggett1985quantum,kofler2007classical}, which are used in other scenarios than Bell's.  

The set Q$^1$ can be described as follows. Let the nonsignaling correlations shared between parties be described by the conditional probability distribution $\Pr(A=k,B=l|\alpha=i,\beta=j)$. At the moment we do not need to specify the finite sets in which the inputs or outputs take their values. Let $\vec{p}_A$ be a (column) vector of marginal probabilities $\Pr(A=k|\alpha=i)$ of Alice, and let $\vec{p}_B$ be the similar vector for Bob. Let the matrix $P$ be the matrix arrangement of probabilities $\Pr(A=k,B=l|\alpha=i,\beta=j)$, such that for every row of $P$ the outcome and input of Alice, i.e., $k$ and $i$ are fixed. Similarly, let the input and output of Bob, $l,j$ be the same for every element on the same column of $P$. The correlations are then said to belong to the set Q$^1$, if the following matrix,
\begin{equation}\label{eq:Gamma}
    \Gamma = \left(
    \begin{array}{c c c}
        1 & \vec{p_A}^T & \vec{p_B}^T\\
        \vec{p_A} & V & P \\
        \vec{p_B} & P^T & W
    \end{array}
    \right),
\end{equation} 
can be made positive semidefinite ($\Gamma\geq 0$) by an appropriate choice of square matrices $V$ and $W$, which themselves are subject to additional constraints:
\begin{equation}
    \mathrm{diag}(V) = \vec{p_A},\; \mathrm{diag}(W) = \vec{p_B}.
\end{equation}
With $\mathrm{diag}(\cdot)$ we denoted a map from the space of $d\times d$ matrices to $\mathds{R}^d$ that takes the diagonal entries of the input matrix.
Additionally, if an off-diagonal entry of $V$ corresponds to the same input of Alice, but two different outputs, then this entry has to be $0$. The same constraints are applied to the off-diagonal entries of $W$ and the inputs and outputs of Bob. A more detailed description of $\Gamma$, in application to our problem, can be found in Appendix~\ref{app:ml}.

We are now ready to present our second main result, which is also stated in terms of the probabilities in Eq.~(\ref{eq:pej}).

\begin{result}\label{th:ml}
The set of constraints from Result~\ref{th:ic} is necessary and sufficient for probabilities $\{p(e|j)\}_{e,j}$ to satisfy the Macroscopic Locality principle.
\end{result}

\begin{proof} 
As introduced earlier in this section, the set of the correlations complying with the ML principle coincides with the set $Q^1$~\cite{navascues2007bounding}, which is defined by the condition $\Gamma \geq 0$ in Eq.~(\ref{eq:Gamma}).
Positivity of $\Gamma$ matrix is in general difficult to resolve for probabilities $\Pr(A=k,B=l|\alpha=i,\beta=j)$ that are not specified. 
More precisely, the condition $\Gamma\geq 0$ is equivalent to a system of polynomial inequalities in $\Pr(A = k,B = l|\alpha = i, \beta = j)$ as well as $V$ and $W$. In order to resolve this system analytically one needs to eliminate the variables $V$
and $W$ from this system. This problem is an instance of the so-called quantifier elimination problems for which the best-known algorithm requires double exponential time~\cite{collins1975quantifier}.
However, in this work we are interested in deriving the set of constraints that the principle of ML sets on the probabilities $p(e|j)$ defined in Eq.~(\ref{eq:pej}). Since the latter forms only a subspace in the space of correlations $\Pr(A=k,B=l|\alpha=i,\beta=j)$, one might hope for simplification of the considered problem. 

Such simplification is indeed possible, if one is able to identify symmetries of the problem, which is a known technique in semidefinite programming~\cite{gatermann2004symmetry,rosset2015characterization}.
According to Ref.~\cite{rosset2015characterization}, if the optimization problem formulated in the hierarchy of Ref.~\cite{navascues2007bounding} exhibits symmetries defined by a group $G$, then the following holds,
\begin{equation}\label{eq:symm_Gamma}
    \Gamma\geq 0 \quad \Leftrightarrow \quad \tilde \Gamma \equiv \frac{1}{|G|}\sum_{g\in G} g.\Gamma \geq 0,
 \end{equation}
 where an action of each group element $g$ on $\Gamma$ corresponds to a permutation of its entries. The implications $\Leftrightarrow$ in Eq.~\ref{eq:symm_Gamma} should be understood in the sense of constraints that the positivity of $\tilde{\Gamma}$ and $\Gamma$ imply on the symmetric subspace, defined by the group $G$, of compatible correlations $\Pr(A = k, B = l| \alpha = i, \beta = j)$.
 As a result of the symmetrization, the number of optimization parameters in $\tilde{\Gamma}$ is reduced compared to the ones in $\Gamma$. Below we identify the symmetry group $G$ in our problem.
 
 First, permutation of Alice's outcome can be counteracted by permutation of Bob's outcome, in a way that the sum of both outcomes remains constant. 
\begin{align}
&\Pr((A\oplus r) \oplus (B \oplus \bar{r}) = i\cdot j\oplus e|\alpha=i,\beta=j)\nonumber\\
=&\Pr(A \oplus B = i\cdot j\oplus e|\alpha=i,\beta=j),
\end{align} 
$\forall r\in [d]$. Second, permutation of Alice's setting can be counteracted by permutation of Bob's outcome, conditioned on Bob's input $j$ in the following way,
\begin{align}
   &\sum_{i=0}^{d-1}\scalebox{0.95}{$\Pr(A\oplus B\oplus s\cdot j=(i\oplus s)j\oplus e|\alpha=i\oplus s,\beta=j)$}\nonumber\\
    &=\sum_{i=0}^{d-1}\Pr(A \oplus B = i\cdot j\oplus e|\alpha=i,\beta=j),
\end{align}
$\forall s\in [d]$. These two types of permutations generate a group $G$, the action of which leaves $p(e|j)$ invariant for all $e$ and $j$. Importantly, the resulting symmetrized matrix $\tilde{\Gamma}$ in Eq.~(\ref{eq:symm_Gamma}) contains only the probabilities $p(e|j)$, constant terms, and optimization variables, number of which is reduced (See Eq.~(\ref{app:eq_tilde_gamma}) in Appendix~\ref{app:ml} for the exact form of $\tilde{\Gamma}$). Note also that we normalize $\tilde{\Gamma}$ by the factor of $d$ in our derivations instead of $|G| = d^2$.

As a next step, we use unitary transformation, $U$ to block-diagonalize the matrix $\tilde{\Gamma}$. In general, diagonalizing such a large matrix is a difficult task, however, $\tilde\Gamma$ itself has an easy-to-handle $d\times d$ block structure (if we do not consider the first row and column), which dictates the form of the unitary $U$, given explicitly in Eq.~(\ref{app:eq_unitary}) in Appendix~\ref{app:ml}. 
  \begin{align}
     \tilde{\Gamma} \stackrel{U}{\longmapsto} \bigoplus_{m=0}^{d+1} \Gamma_m,
  \end{align}
 where $\Gamma_0$ is a $(d+1)\times (d+1)$ constant positive matrix, independent of $p(e|j)$ (see Eq.~(\ref{app:eq_gamma_0}) in Appendix~\ref{app:ml}) and $\Gamma_m$, for $1 \leq m\leq d+1$, are $d\times d$ Hermitian matrices. To be concise, we only give the upper-triangular part of $\Gamma_m$ here,
 
 \begin{equation}\label{eq:Gamma_m}
     \Gamma_{m}=\begin{pmatrix}1 & \nu_{m}^{1} & \dots & \nu_{m}^{d-1} & p_{m}^{0} & p_{m}^{1}\\
 & 1 & \ddots & \vdots & p_{m}^{0} & \omega^{m}p_{m}^{1}\\
 &  & \ddots & \nu_{m}^{1} & \vdots & \vdots\\
 &  &  & 1 & p_{m}^{0} & \omega^{(d-1)\cdot m}p_{m}^{1}\\
 &  &  &  & 1 & 0\\
 &  &  &  &  & 1
\end{pmatrix},
 \end{equation}
where $p^j_m = \sum_{e=0}^{d-1}\omega^{m\cdot(1+e)}p(e|j)$ and $v_m^i$ are optimization variables. 

We have, thus, reduced checking compatibility of $p(e|j)$ with the ML principle to positivity of block-diagonal matrices: 
\begin{equation}
    \Gamma \geq 0 \quad \Leftrightarrow \quad \Gamma_m\geq 0, \quad 1\leq m\leq d+1. 
\end{equation}
Now we can finally show that the inequalities in Eq.~(\ref{eq:main}) derived from IC principle, are necessary and sufficient for  $ \Gamma \geq 0$.
First, we prove that if $\Gamma_m\geq 0$, $\forall m$, then the condition in Eq.~(\ref{eq:main}) must hold.  If we use the notation, where $V_m = \sum_{i,i'=0}^{d-1}\nu^{i\ominus i'}_m\kb{i}{i'}$, with the convention that $\nu^0_m = 1$ and $\nu^{i-i'}_m = (\nu^{i'-i}_m)^*$, and $P_m$ is a $d\times 2$ matrix with the first column given by $p^0_m$ and the second by the vector $\sum_{i=0}^{d-1}\omega^{m\cdot i}p^1_m\ket{i}$, we can formulate the positivity of $\Gamma_m$ in terms of the Schur complement:
 \begin{equation}
    \Gamma_m=\begin{pmatrix}
    V_m & P_m \\
    P_m^\dagger & \openone
    \end{pmatrix} \geq 0 \; \Leftrightarrow \;  V_m-P_m P_m^\dagger \geq 0.
\end{equation}
 The latter condition implies $1-|p_m^0|^2-|p_m^1|^2\geq 0$ from the positivity of the diagonal elements. Inserting the values of $p_m^0$ and $p_m^1$ for $1\leq m\leq d-1$, we get the familiar expression, 
\begin{align}
  \left|\sum_{e=0}^{d-1}p(e|0)\omega^{e \cdot m}\right|^2+\left|\sum_{e=0}^{d-1}p(e|1)\omega^{e \cdot m}\right|^2\leq 1.
 \end{align}
 However, since these inequalities are symmetric under conjugation, we can remove redundant ones by just considering $m\in\{1,\dots \left\lfloor \frac{d}{2}\right\rfloor \}$. This finishes one direction of the proof. 
 
 For the reverse direction, we need to show that if inequalities in Eq.(\ref{eq:main}) hold, then we can always find $v^i_m$ in Eq.~(\ref{eq:Gamma_m}), such that $\Gamma_m \geq 0$, or alternatively:
 \begin{align}\label{eq:reversedirectionMLVM}
     V_m-P_mP_m^\dagger\geq 0.
 \end{align}
 If we choose the variables of $V_m$ to be equal to the off-diagonal entries of $P_mP_m^\dagger$, 
 \begin{equation}\label{eq:A_coeff}
     \nu_m^i=|p_m^0|^2+|p_m^1|^2 \omega^{-i \cdot m},
 \end{equation}
 then the condition $\Gamma_m \geq 0$ indeed follows from $|p_m^0|^2+|p_m^1|^2\leq 1$. 
  \end{proof} 
  
Combining the two main results of this work, Results~\ref{th:ic} and~\ref{th:ml}, we can directly conclude that in the entire subspace of probabilities $p(e|j)$, the principle of IC provides at least as good approximation of the set of quantum correlations as the ML principle. On the other hand, results of Refs.~\cite{cavalcanti2010macroscopically,miklin2021information} showed that there exist macroscopically local correlations $\{p(e|j)\}_{e,j}$ that violate the IC principle. Hence, we can conclude the following:

\begin{result}
In the subspace of correlations, defined by the probabilities $\{p(e|j)\}_{e,j}$, the principle of Information Causality is strictly stronger than that of Macroscopic Locality.
\end{result}

\begin{figure}	
\centering
	\includegraphics[width=0.3\textwidth]{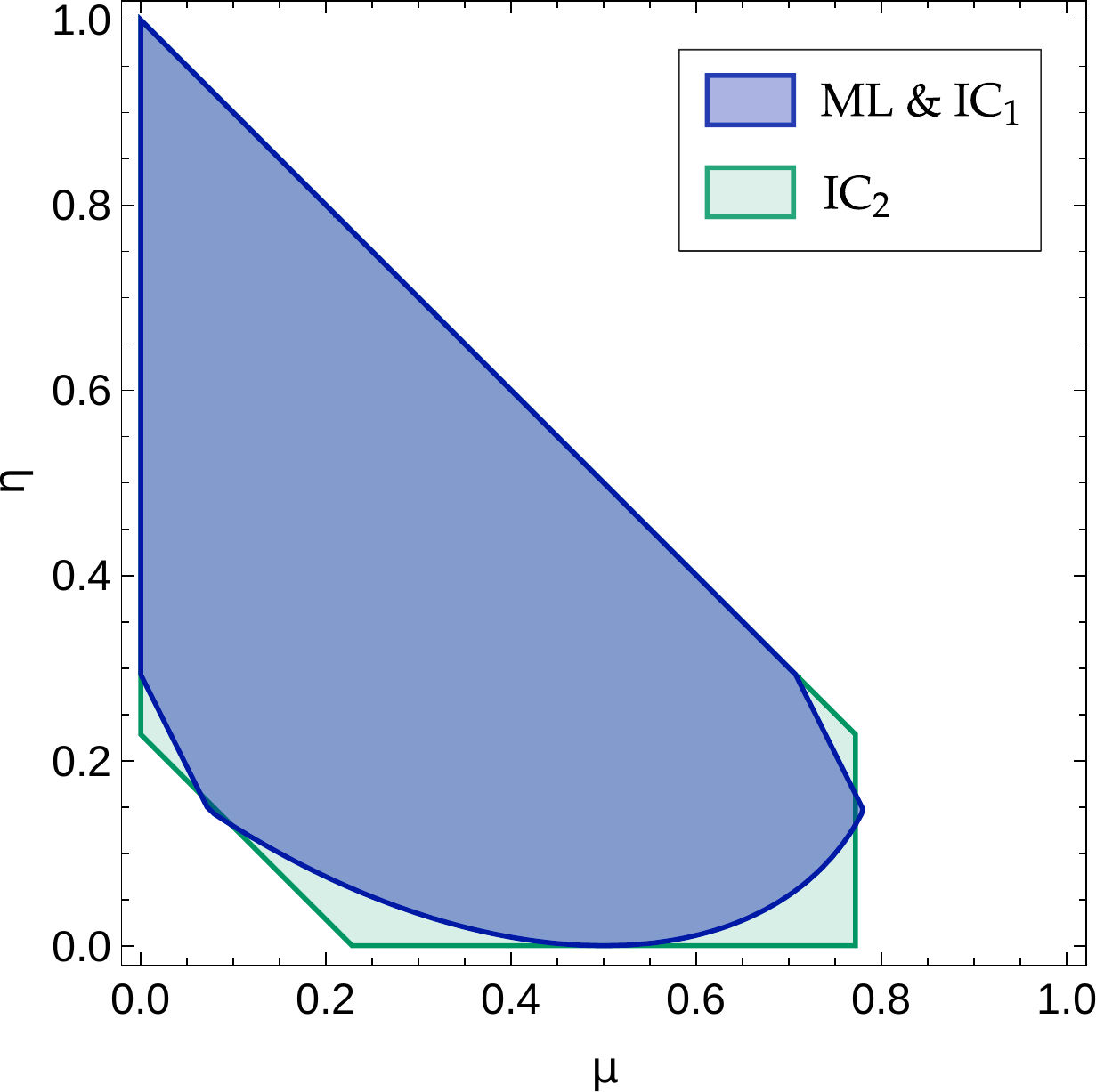} 
\caption{The regions constrained by the IC and ML principles for $p(0|0)=\mu, p(1|0)=p(2|0)=\frac{\eta}{2}$, and $p(e|0)=p(e|1)$ for all $e\in[d]$, where $d=4$. ML $\&$ IC$_1$ (dark blue) is the set constrained by both the IC and ML principles as follows from Result~\ref{th:ic}. IC$_2$ (light green) is an alternative approximation of the set constrained by IC taken from Ref.~\cite{miklin2021information}.}
\label{fig:2}
\end{figure}

In Fig.~\ref{fig:2} we give a simple example of an inclusion of the region of correlations constrained by IC in that of ML. We consider the assignment of probabilities $p(0|0)=\mu, p(1|0)=p(2|0)=\frac{\eta}{2}$, and $p(e|0)=p(e|1)$ for all $e\in[d]$, where $d=4$. The set of correlations constrained by IC is included in the intersection of the sets depicted by ML $\&$ IC$_1$ and IC$_2$ in Fig.~\ref{fig:2}, while the correlations obeying the ML principle are exactly those denoted by ML $\&$ IC$_1$ in Fig.~\ref{fig:2}. One can notice two small regions that are included in ML $\&$ IC$_1$ but excluded from IC$_2$. Examples of the parameter values belonging to these regions are $\mu = \frac{2+3\sqrt{2}}{8}$, $\eta = \frac{2-\sqrt{2}}{4}$ and $\mu = \frac{2-\sqrt{2}}{8}$, $\eta = \frac{2-\sqrt{2}}{4}$.

Here IC$_2$ is an alternative approximation of the set constrained by the IC principle taken from Ref.~\cite{miklin2021information}. In Ref.~\cite{miklin2021information} one is only concerned with optimally bounding the ``guessing'' probabilities $p(0|0)$ and $p(0|1)$, rather than aiming at a tight approximation of the whole set of probabilities. To that end, Ref.~\cite{miklin2021information} uses Fano's inequality~\cite{fano1968transmission} instead of Eq.~(\ref{eq:fano}). For further information on this approach, see Discussions.

\section{Discussions}
In this work we consider the problem of bounding the set of quantum correlations, observed in a Bell test, in the space of all nonsignaling correlations. To this end, we provide a family of quantum Bell inequalities. We derive our inequalities from the principle of Information Causality and show that in the subspace of probabilities, defined by symmetries of these inequalities, they provide necessary and sufficient conditions for the principle of Macroscopic Locality. From these two results we conclude that in the aforementioned subspace the principle of Information Causality is strictly stronger than that of Macroscopic Locality. 

Although our main results are tailored to the scenario in which Bob has binary settings, the derived inequalities can be applied to arbitrary Bell scenarios. For instance, one can adjust our inequalities to the scenarios with lower number of outcomes by simply setting the probabilities of some outcomes to $0$. On the other hand, by using the derived inequalities for each pair of Bob's inputs, one can bound correlations in the scenarios with an arbitrary number of Bob's settings.

The derivation of our main result uses concatenation protocol, while in the recent paper of Ref.~\cite{miklin2021information} it was claimed that all known results can be re-derived and improved by replacing this protocol with communication over a noisy classical channel. A natural question to ask is whether concatenation is really necessary here, or could we have gone without it. It happens to be the former since the results obtained in the current work differ qualitatively from all the bounds published before and referenced in Ref.~\cite{miklin2021information}. The idea of Ref.~\cite{miklin2021information} is to look at the concatenation procedure as a communication channel taking the outputs of the topmost ($(n-1)$-th) layer on Alice's side and producing the message that Bob uses with his final box to obtain his guess. The transition matrix of this channel is given by the probability distribution of Bob's guesses given Alice's data. If a protocol involving concatenation procedure leads to a violation of Information Causality, then there exists a protocol, which also gives a violation, but involves only a single box and a communication over a channel described by this transition matrix. However, Ref.~\cite{miklin2021information} uses an implicit assumption that the transition matrix is the same for all Alice's dits. This is the case for Refs.~\cite{pawlowski2009information,pawlowski2010entanglement,cavalcanti2010macroscopically}  but not for the general protocol studied in the current work.

Our results can be generalized in many directions. One could use the techniques of this paper to derive inequalities for the scenario in which Bob has ternary inputs by extending the concatenation protocol from Ref.~\cite{pawlowski2010entanglement}. The techniques used in the proof regarding ML principle can be employed to derive quantum Bell inequalities displaying other symmetries. Moreover, a tighter approximation of the entropy function will give tighter inequalities, which might shed light on the long-standing problem of whether almost quantum correlations can violate the IC principle~\cite{navascues2015almost}. 

Resolving this problem is hard for a few reasons. First, there are numerous ways to employ nonsignaling correlations in the communication scenario of the IC, which makes it difficult to decide whether a given nonsignaling box violates the IC. Second, the set of almost quantum correlations is believed to give a tight approximation to the quantum set, allowing for no room for relaxations of the IC statement. In this case, one might need to consider combinations of various nonsignaling boxes in the same communication protocol. We believe that the approach of deriving quantum Bell inequalities is advantageous over testing individual candidates of nonsignaling boxes as the former is more systematic for high-dimensional correlations spaces.

\section*{Acknowledgments}
\vspace{-0.3cm}
We thank Costantino Budroni and David Gross for fruitful discussions.  We acknowledge partial support by the Foundation for Polish Science (IRAP project, ICTQT, contract no. 2018/MAB/5, co-financed by EU within Smart Growth Operational Programme). This research was made possible by funding from QuantERA, an ERA-Net cofund in Quantum Technologies (www.quantera.eu), under the project eDICT. M.P.~acknowledges the support of NCN through grant SHENG (2018/30/Q/ST2/00625). Funded by the Deutsche Forschungsgemeinschaft (DFG, German Research Foundation) under Germany's Excellence Strategy – Cluster of Excellence Matter and Light for Quantum Computing (ML4Q) EXC 2004/1 – 390534769

\newpage
\onecolumngrid
\begin{appendix}
\section*{APPENDIX}
In this Appendix, we provide technical details regarding proofs and derivations of the results stated in the main text.

One of the most useful identities, which we use in the proofs, are the sums related to Dirichlet kernel~\cite{lejeune1829convergence}. See Ref.~\cite{gradshtein2007table} (section 1.341) for more exhaustive list. 
\begin{align}
\sum_{k=1}^{n-1}\sin(x+ky) = & \sin{\left(x+\frac{n-1}{2}y\right)}\sin\left({\frac{ny}{2}}\right)\csc\left({\frac{y}{2}}\right). \label{eq:dirichlet1}\\
\sum_{k=1}^{n-1}\cos(x+ky) = & \cos{\left(x+\frac{n-1}{2}y\right)}\sin\left({\frac{ny}{2}}\right)\csc\left({\frac{y}{2}}\right).\label{eq:dirichlet2}\\
\sum_{k=1}^{n-1}(-1)^k\sin(x+ky) = & \sin{\left(x+\frac{n-1}{2}(y+\pi)\right)}\sin\left({\frac{n(y+\pi)}{2}}\right)\sec\left({\frac{y}{2}}\right).\label{eq:dirichlet3}\\
\sum_{k=1}^{2n-1}(-1)^k\cos(x+ky) = & \sin{\left(x+\frac{2n-1}{2}y\right)}\sin({ny})\sec\left({\frac{y}{2}}\right)\label{eq:dirichlet4}.
\end{align}
We refer to these identities in several parts of this Appendix.

We also use the following notation in this Appendix: $\mathbbm{J}$ is $d\times d$ matrix of all $1$'s, i.e., $\mathbbm{J} = \sum_{i,j=0}^{d-1}\kb{i}{j}$. The matrix $S_E=\sum_{i=0}^{d-1}\ket{i}\bra{\bar{i}\oplus E}$. Generalized $d$-dimensional Pauli matrices are defined as follows,
\begin{equation}
X^{m}=~\sum_{i=0}^{d-1}\ket{i}\bra{i\ominus m} \quad \mbox{ and } \quad Z^{m}=~\sum_{i=0}^{d-1}\omega^{i \cdot m}\ket{i}\bra{i}.    
\end{equation}
Generalized $d$-dimensional Hadamard is given by 
\begin{equation}
    H=\frac{1}{\sqrt{d}}\sum_{i,j=0}^{d-1}\omega^{- i 
    \cdot j}\ket{i}\bra{j}.
\end{equation}
Finally, the following identities will prove useful,
\begin{equation}
    H^{\dagger}X^m H =Z^m, \qquad S_{i}S_{d-1}=X^{i+1} \quad \mbox{ and } \quad S_{d-1}^\dagger X^m S_{d-1}=X^{-m}.
\end{equation}

\section{Proof of Result~\ref{th:ic}}
\subsection{Proof of Eq.~(\ref{eq:Q_odd}) and the even case}\label{app:Q_induction}
Here we prove the formula for $\ket{Q^{(k)}(j)}$ given in  Eq.~(\ref{eq:Q_odd}) for $d$ odd and its even counterpart: 
\begin{align}
    Q^{(k)}_{e}(j) &= \frac{1}{d}+\frac{1}{d}\sum_{i=1}^{\frac{d-1}{2}}{\left(l^k_i(j)\omega^{e\cdot i}+{l^*_i}^k(j)\omega^{-e\cdot i}\right)},\quad \forall e\in [d],\quad  d\;\text{odd},\label{eq:app_q_odd}\\
    Q^{(k)}_{e}(j) &= \frac{1}{d}+\frac{1}{d}\sum_{i=1}^{\frac{d-2}{2}}{\left(l^k_i(j)\omega^{e\cdot i}+{l^*_i}^k(j)\omega^{-e\cdot i}\right)}+\frac{1}{d}l^k_{\frac{d}{2}}(j)(-1)^e,\quad \forall e\in [d],\quad  d\;\text{even},\label{eq:app_q_even}
\end{align}
where 
\begin{equation}\label{eq:app_l}
    l_i(j) = \sum_{e=0}^{d-1}p(e|j)\omega^{-e\cdot i},\; i \in \left\{1,\dots,\left\lfloor\frac{d}{2}\right\rfloor\right\}.
\end{equation}
\begin{proof}
We split the proof in two parts, first we prove the above expressions for $d$ odd, and then for $d$ even. It is clear that the proof is independent of the choice of Bob's setting $j$. 

\textbf{For odd} $d$, we start by proving the base case of $k=1$. 
\begin{align}\label{eq:app_proof_q_base_odd}
   \frac{1}{d}+\frac{1}{d}\sum_{i=1}^{\frac{d-1}{2}}\sum_{e'=0}^{d-1}p(e'|j)\left(\omega^{(e-e')\cdot i}+\omega^{(e'-e)\cdot i}\right)= \frac{1}{d}+\frac{1}{d}\sum_{e'=0}^{d-1}p(e'|j)\cdot 2\sum_{i=1}^{\frac{d-1}{2}}\cos\left(\frac{2\pi(e-e')\cdot i}{d}\right).
\end{align}
We separately write out the last sum and use Eq.~(\ref{eq:dirichlet2}) to evaluate it,
\begin{equation}
    2\sum_{i=1}^{\frac{d-1}{2}}\cos\left(\frac{2\pi(e-e')\cdot i}{d}\right) = \frac{\sin\left(\frac{2\pi(e-e')d}{2d}\right)}{\sin\left(\frac{2\pi(e-e')}{2d}\right)}-1.
\end{equation}
In this expression, if $e=e'$, the sum amounts to $d-1$, and if $e\neq e'$, to $-1$. Inserting these two cases in Eq.~(\ref{eq:app_proof_q_base_odd}), we prove the base case $Q^{(1)}_e(j)=p(e|j)$,
\begin{align}
    \frac{1}{d}-\frac{1}{d}(1-p(e|j))+\frac{1}{d}p(e|j)(d-1)=p(e|j).
\end{align}

Now we can prove the induction step by assuming the form of $Q^{(k)}_e(j)$ in Eq.~(\ref{eq:app_q_odd}) for some $k$, and using the general formula $\ket{Q^{(k+1)}(j)} = M_j\ket{Q^{(k)}(j)}$ to calculate the $(k+1)$ case. 
\begin{align}\label{eq:app_q_induction_step_odd}
\begin{split}
    Q^{(k+1)}_e(j) &= \sum_{e'=0}^{d-1}Q^{(k)}_{e'}(j)p(e\oplus \bar{e'}|j) = \frac{1}{d}+\frac{1}{d}\sum_{i=1}^{\frac{d-1}{2}}l^k_i(j)\left(\sum_{e'=0}^{d-1}\omega^{e'\cdot i}p(e\oplus \bar{e'}|j)\right)\\
    &+\frac{1}{d}\sum_{i=1}^{\frac{d-1}{2}}{l^*_i}^k(j)\left(\sum_{e'=0}^{d-1}\omega^{-e'\cdot i}p(e\oplus \bar{e'}|j)\right).
    \end{split}
\end{align}
It should be clear that if we take factors $\omega^{e\cdot i}$ and $\omega^{-e\cdot i}$ out of the sums over $e'$ above, these sums will become $l_i(j)$ and $l^*_i(j)$ respectively, since in the definition of $l_i(j)$ in Eq.~(\ref{eq:app_l}) we can always shift the index $e$ of the sum by some constant $e'$.

\textbf{For even} $d$, we repeat the procedure above and start with the base case of $k=1$. This time we need to show that
\begin{align}\label{eq:app_proof_q_base_even}
    \frac{1}{d}+\frac{1}{d}\sum_{i=1}^{\frac{d-2}{2}}\sum_{e'=0}^{d-1}p(e'|j)\left(\omega^{(e-e')\cdot i}+\omega^{(e'-e)\cdot i}\right)+\frac{1}{d}\sum_{e'=0}^{d-1}(-1)^{(e-e')}p(e'|j) = p(e|j).
\end{align}
We again use the Dirichlet identity from Eq.~(\ref{eq:dirichlet2}) to calculate the sum of $\omega^{(e-e')\cdot i}+\omega^{(e'-e)\cdot i}$ over $i$ and $e'$.
\begin{equation}\label{eq:app_dirichlet_even}
        2\sum_{i=1}^{\frac{d-2}{2}}\cos\left(\frac{2\pi(e-e')\cdot i}{d}\right) = \frac{\sin\left(\frac{2\pi(e-e')(d-1)}{2d}\right)}{\sin\left(\frac{2\pi(e-e')}{2d}\right)}-1 = \frac{\sin\left(\pi(e-e')-\frac{\pi(e-e')}{d}\right)}{\sin\left(\frac{\pi(e-e')}{d}\right)}-1.
\end{equation}
This time the conclusions are different. If $e=e'$, the sum is equal to $d-2$. If $(e-e')$ is even (except for $0$), then the expression in Eq.~(\ref{eq:app_dirichlet_even}) is equal to $-2$. If, on the other hand, $(e-e')$ is odd, the expression is equal to $0$. We can express the last findings in the following way:
\begin{equation}
    \sum_{i=1}^{\frac{d-2}{2}}\sum_{e'=0}^{d-1}p(e'|j)\left(\omega^{(e-e')\cdot i}+\omega^{(e'-e)\cdot i}\right) = (d-2)p(e|j)-\sum_{e'\neq e}(1+(-1)^{(e-e')})p(e'|j).
\end{equation}
The last sum on the left-hand side of Eq.~(\ref{eq:app_proof_q_base_even}) separates into the term $p(e|j)$, corresponding to $e=e'$ and the rest, which is $\sum_{e'\neq e}(-1)^{(e-e')}p(e'|j)$. It is clear then, that summing all the terms up, leads to the proof of Eq.~(\ref{eq:app_proof_q_base_even}). 
Now, the missing part of the proof is the induction step for the case of even $d$. We again assume the form of $Q^{(k)}_e(j)$ and explicitly write the $k+1$ case as in Eq.~(\ref{eq:app_q_induction_step_odd}).
\begin{align}
\begin{split}
    Q^{(k+1)}_e(j) &= \frac{1}{d}+\frac{1}{d}\sum_{i=1}^{\frac{d-2}{2}}l^k_i(j)\left(\sum_{e'=0}^{d-1}\omega^{e'\cdot i}p(e\oplus \bar{e'}|j)\right)+\frac{1}{d}\sum_{i=1}^{\frac{d-2}{2}}{l^*_i}^k(j)\left(\sum_{e'=0}^{d-1}\omega^{-e'\cdot i}p(e\oplus \bar{e'}|j)\right)\\
    &+\frac{1}{d}l^k_{\frac{d}{2}}\sum_{e'=0}^{d-1}(-1)^{e'}p(e\oplus \bar{e'}|j).
    \end{split}
\end{align}
As in the case of odd $d$, one can easily see by taking the factors of $\omega^{e\cdot i}$, $\omega^{-e\cdot i}$, and $(-1)^{e}$ out of the sums over $e'$ that these sums are equal to $l_i(j)$, $l^*_i(j)$, and $l_{\frac{d}{2}}(j)$ respectively. 
\end{proof}

\subsection{Proof of Eq.~(\ref{eq:probderivations3})}\label{app:proofOfMain}
Here we rigorously derive Eq.~(\ref{eq:probderivations3}) as well as give the similar expression for even $d$. We start with \textbf{odd} $d$. 
\begin{align}
\begin{split}
        P(E|k)-\frac{1}{d}=&\sum_{e=0}^{d-1}Q^k_{e}(0)Q^{n-k}_{\bar e\oplus E}(1)-\frac{1}{d}\\
        =& \frac{1}{d^2} \sum_{i_1,i_2=1}^{\frac{d-1}{2}}\sum_{e=0}^{d-1}{\left(\omega^{i_1 e}l_{i_1}^k(0)+\omega^{-i_1 e}l_{i_1}^{* k}(0)\right)}{\left(\omega^{i_2 (\bar e\oplus E)}l_{i_2}^{n-k}(1)+\omega^{-i_2 (\bar e\oplus E)}l_{i_2}^{*(n-k)}(1)\right)}\\
  =&\frac{4}{d^2}\sum_{i_1,i_2=1}^{\frac{d-1}{2}}\sum_{e=0}^{d-1}\left(\cos\left(\frac{2\pi e i_1}{d}\right)m^{(k)}_{i_1}-\sin\left(\frac{2\pi e i_1}{d}\right)n^{(k)}_{i_1}\right) \\
  &\cdot\left(\cos\left(\frac{2\pi (\bar e\oplus E) i_2}{d}\right)m^{(n-k)}_{i_2}-\sin\left(\frac{2\pi (\bar e\oplus E) i_2}{d}\right)n^{(n-k)}_{i_2}\right)\label{app:eq_proofMain3}.
\end{split}
\end{align}
In Eq.~(\ref{app:eq_proofMain3}) the expressions $m_{i_1}^{(k)}$ and $n_{i_1}^{(k)}$ denote the real and imaginary parts of $l_{i1}^k(0)$.  In the same way, we denote the real and imaginary parts of $l_{i2}^{n-k}(1)$ by $m_{i_2}^{(n-k)}$ and $n_{i_2}^{(n-k)}$, respectively. First we take a sum over $e$. Note that, conveniently, all $l_i(j)$ are constant in $e$. 
\begin{align}
    \sum_{e=0}^{d-1}\cos\left(\frac{2\pi e i_1}{d}\right)\cos\left(\frac{2\pi (\bar e\oplus E) i_2}{d}\right)&=\frac{d}{2}\cos\left(\frac{2\pi E i}{d}\right),\\
    \sum_{e=0}^{d-1}\cos\left(\frac{2\pi e i_1}{d}\right)\sin\left(\frac{2\pi (\bar e\oplus E) i_2}{d}\right)&= \frac{d}{2}\sin\left(\frac{2\pi E i}{d}\right),\\
    \sum_{e=0}^{d-1}\sin\left(\frac{2\pi e i_1}{d}\right)\cos\left(\frac{2\pi (\bar e\oplus E) i_2}{d}\right)&=\frac{d}{2}\sin\left(\frac{2\pi E i}{d}\right),\\
    \sum_{e=0}^{d-1}\sin\left(\frac{2\pi e i_1}{d}\right)\sin\left(\frac{2\pi (\bar e\oplus E) i_2}{d}\right)
    &=-\frac{d}{2}\cos\left(\frac{2\pi E i}{d}\right),
\end{align}
where $i=i_1=i_2$ (See identities in Eqs.~(\ref{eq:dirichlet1},\ref{eq:dirichlet2})).
Putting all the values in Eq.~(\ref{app:eq_proofMain3}) we get,
\begin{align}
\begin{split}
        P(E|k)-\frac{1}{d}=&\frac{2}{d}\sum_{i=1}^{\frac{d-1}{2}}\left(\cos\left({\frac{2 \pi E i}{d}}\right)(m_i^{(k)}m_i^{(n-k)}-n_i^{(k)}n_i^{(n-k)})-\sin\left({\frac{2 \pi E i}{d}}\right)(m_i^{(k)}n_i^{(n-k)}+n_i^{(k)}m_i^{(n-k)})\right)\\
        =&\frac{2}{d}\sum_{i=1}^{\frac{d-1}{2}}\left(\cos\left({\frac{2 \pi E i}{d}}\right)A_i-\sin\left({\frac{2 \pi E i}{d}}\right)B_i\right).
\end{split}
\end{align}
\textbf{Even} $d$: 
\begin{align}
        P(E|k)- \frac{1}{d}&= \sum_{e=0}^{d-1}Q^k_{e}(0)Q^{n-k}_{\bar e\oplus E}(1)-\frac{1}{d}=\frac{(-1)^E}{d}l_{\frac{d}{2}}^k(0)l_{\frac{d}{2}}^{n-k}(1)\nonumber\\
  &+\frac{4}{d^2}\sum_{i_1,i_2=1}^{\frac{d-2}{2}}\sum_{e=0}^{d-1}\left(\cos\left(\frac{2\pi e i_1}{d}\right)m^{(k)}_{i_1}-\sin\left(\frac{2\pi e i_1}{d}\right)n^{(k)}_{i_1}\right) \nonumber \\
  &\cdot \left(\cos\left(\frac{2\pi (\bar e\oplus E) i_2}{d}\right)m^{(n-k)}_{i_2}-\sin\left(\frac{2\pi (\bar e\oplus E) i_2}{d}\right)n^{(n-k)}_{i_2}\right)\nonumber\\
  &+ \frac{2}{d^2}\sum_{i_1,i_2=1}^{\frac{d-2}{2}}\sum_{e=0}^{d-1}(-1)^{\bar{e}\oplus E} \left(\cos\left(\frac{2\pi e i_1}{d}\right)m^{(k)}_{i_1}-\sin\left(\frac{2\pi e i_1}{d}\right)n^{(k)}_{i_1}\right)l_{\frac{d}{2}}^{n-k}(1)\label{app:eq_extraterm1}\\
  &+ \frac{2}{d^2}\sum_{i_1,i_2=1}^{\frac{d-2}{2}}\sum_{e=0}^{d-1}(-1)^e l_{\frac{d}{2}}^{k}(0)\left(\cos\left(\frac{2\pi (\bar e\oplus E) i_2}{d}\right)m^{(n-k)}_{i_2}-\sin\left(\frac{2\pi (\bar e\oplus E) i_2}{d}\right)n^{(n-k)}_{i_2}\right)\label{app:eq_extraterm2}.
\end{align}
We first show that the summands in Eqs.~(\ref{app:eq_extraterm1},\ref{app:eq_extraterm2}) amount to $0$. Using Eqs.~(\ref{eq:dirichlet3},\ref{eq:dirichlet4}), we show that
\begin{align}
    \sum_{e=0}^{d-1}(-1)^{\bar{e}}\cos\left(\frac{2\pi e i_1}{d}\right)= & \sin\left(\frac{(d-1)\pi i_1}{d}\right)\sin(\pi i_1)\sec\left({\frac{\pi i_1}{d}}\right)=0.\\
    \sum_{e=0}^{d-1}(-1)^{\bar{e}}\sin\left(\frac{2\pi e i_1}{d}\right)= &\sin\left(\frac{(d-1)(2\pi i_1+d \pi)}{2d}\right)\sin\left(\pi (i_1+\frac{d}{2})\right)\sec\left({\frac{\pi i_1}{d}}\right)=0.
\end{align}
Similarly, the summand in Eq.~(\ref{app:eq_extraterm2}) amount to $0$ too. Hence, the expression for $P(E|k)$ becomes
\begin{align}\label{eq:app_evendProbabilities}
        P(E|k)-\frac{1}{d}=\frac{2}{d}\sum_{i=1}^{\frac{d-2}{2}}\left(\cos\left({\frac{2 \pi E i}{d}}\right)A_i-\sin\left({\frac{2 \pi E i}{d}}\right)B_i\right)+\frac{(-1)^E}{d}l_{\frac{d}{2}}^k(0)l_{\frac{d}{2}}^{n-k}(1).
\end{align}

\subsection{Proof of inequality in Eq.~(\ref{eq:fano})}\label{app:fano}
First, we repeat claim: 
For two random variables $a$ and $b$ taking values in $[d]$, and $a$ being uniformly distributed, define variable $e$, such that $\Pr(b=a\oplus e)=1$. Then, the following inequality holds:
\begin{equation}
    I(a;b)\geq \ln d - H(e).
\end{equation}
\begin{proof}
Let $\{p(a=i,b=j)\}_{i,j}$ be a joint probability distribution of $a$ and $b$ and $\{p_a(a=i)\}_i$ and $\{p_b(b=j)\}_j$ the corresponding marginal distributions of $a$ and $b$ respectively. In the following we use the shorthand notation $p(i,j)\equiv p(a=i,b=j)$, and $p_a(i)\equiv p_a(a=i)$, $p_b(j)\equiv p_b(b=j)$. Since $a$ is uniformly distributed, $p_a(i)=\frac{1}{d}, \forall i\in [d]$. We start by writing explicitly the form of $I(a;b)$:
\begin{equation}
    I(a;b) = \sum_{i=0}^{d-1}\sum_{j=0}^{d-1}p(i,j)\ln \frac{p(i,j)}{p_a(i)p_b(j)}=\sum_{k=0}^{d-1}\sum_{i=0}^{d-1}p(i,i\oplus k) \ln \frac{p(i,i\oplus k)}{p_a(i)p_b(i\oplus k)},
\end{equation}
Now we use the log sum inequality (see e.g., Ref.~\cite{cover1999elements}) for $c_i,d_i\in \mathds{R}$:
\begin{equation}
    \sum_{i}c_i \ln \frac{c_i}{d_i}\geq \left(\sum_i c_i\right)\ln \frac{\sum_i c_i}{\sum_i d_i },
\end{equation}
to obtain the following:
\begin{equation}
    I(a;b)\geq \sum_{k=0}^{d-1}\left(\sum_{i=0}^{d-1}p(i,i\oplus k)\right) \ln \frac{\sum_{i=0}^{d-1}p(i,i\oplus k)}{\sum_{i=0}^{d-1} p_a(i)p_b(i\oplus k)} = \sum_{k=0}^{d-1}{p}(e=k)\ln \frac{{p}(e=k)}{\frac{1}{d}},
\end{equation}
where we used the fact that $a$ is uniformly distributed. The probabilities ${p}(e=k) \equiv \sum_{i=0}^{d-1}p(i,i\oplus k)$ obviously form a distribution for $e$, corresponding to $\{p(i,j)\}_{i,j}$. Hence, $H(e) = - \sum_{k=0}^{d-1}{p}(e=k)\ln {p}(e=k)$, which proves the inequality.
\end{proof}

\subsection{Approximation of entropy with polynomial of degree 2}\label{app:approx_entr}
Given a random variable $a$ taking values in $[d]$, and $\{p(a=i)\}_i$ a distribution of $a$, there exist $\kappa>0$, such that the following inequality holds:
\begin{equation}
   H(a) = -\sum_{i=0}^{d-1}p(a=i)\ln p(a=i) \leq \ln d - \kappa \sum_{i=0}^{d-1}\left( p(a=i)-\frac{1}{d}\right)^2,
\end{equation}
where it is sufficient to take $\kappa = \frac{d^2(\ln d-1)+d}{(d-1)^2}$.
\begin{proof}
Let $x_i = dp(a=i)-1, \forall i\in [d]$. Let us write the expression for $p(a=i)\ln p(a=i)$ in terms of $x_i$:
\begin{equation}\label{eq:app_entr_approx_kappa}
    p(a=i)\ln p(a=i) = \left(\frac{x_i}{d}+\frac{1}{d}\right)\ln \left(\frac{x_i}{d}+\frac{1}{d}\right) = \frac{1}{d}(x_i+1)(-\ln d+\ln(x_i+1)).
\end{equation}
Since $\sum_{i=0}^{d-1}p(a=i)=1$, we have that $\sum_{i=0}^{d-1}x_i = 0$. Also, due to positivity of probability, we have that $-1\leq x_i \leq d-1, \forall i\in[d]$. On the other hand, the function $(x_i+1)\ln (x_i+1)-x_i$ behaves like $\frac{x_i^2}{2}$ around $x_i=0$. Hence, we can find $\kappa'$, such that $(x_i+1)\ln (x_i+1)-x_i\geq \kappa' x_i^2$ for the domain of $x_i$. The sufficient coefficient $\kappa'$ can be determined by taking $x_i = (d-1)$, which gives:
\begin{equation}
    \kappa' = \frac{d\ln d-d+1}{(d-1)^2}.
\end{equation}
We finish the proof by substituting this bound in Eq.~(\ref{eq:app_entr_approx_kappa}) and summing over $i$ (and remembering that $\sum_{i=0}^{d-1}x_i = 0$):
\begin{align}\begin{split}
    \sum_{i=0}^{d-1}p(a=i)\ln p(a=i)&\geq -\ln d+\frac{1}{d}(1-\ln d)\sum_{i=0}^{d-1}x_i+\frac{k'}{d}\sum_{i=0}^{d-1}x_i^2\\
    &\geq -\ln d+\frac{d^2(\ln d-1)+d}{(d-1)^2}\sum_{i=0}^{d-1}\left(p(a=i)-\frac{1}{d}\right)^2.
    \end{split}
\end{align}
\end{proof}

\subsection{Proof of Eq.~(\ref{eq:theorem1prooffinalStep})}\label{App:theorem1prooffinalStep}
The techniques employed in this proof are very similar to the ones used in Eq.~(\ref{eq:probderivations3}) and rely on trigonometric identities in Eq.~(\ref{eq:dirichlet1},\ref{eq:dirichlet2}). We consider the simpler case first, the case of \textbf{odd} $d$.

\begin{align}
       \sum_{E=0}^{d-1} \left(P(E|k)-\frac{1}{d}\right)^2=& \frac{4}{d^2}\sum_{i_1,i_2=1}^{\frac{d-1}{2}}      \sum_{E=0}^{d-1}\left(\cos\left({\frac{2 \pi E i_1}{d}}\right)A_{i_1}-\sin\left({\frac{2 \pi E i_1}{d}}\right)B_{i_1}\right)\nonumber \\
       &\cdot \left(\cos\left({\frac{2 \pi E i_2}{d}}\right)A_{i_2}-\sin\left({\frac{2 \pi E i_2}{d}}\right)B_{i_2}\right)\nonumber\\
      =&\frac{2}{d}\sum_{i=1}^{\frac{d-1}{2}}\left(A_i^2+B_i^2\right)\\ 
      =&\frac{2}{d}\sum_{i=1}^{\frac{d-1}{2}}\left((m_i^{(k)}m_i^{(n-k)})^2+(n_i^{(k)}n_i^{(n-k)})^2+(m_i^{(k)}n_i^{(n-k)})^2+(n_i^{(k)}m_i^{(n-k)})^2\right)\nonumber \\
      =&\frac{2}{d}\sum_{i=1}^{\frac{d-1}{2}}\left((m_i^{(k)})^2+(n_i^{(k)})^2\right)\left((m_i^{(n-k)})^2+(n_i^{(n-k)})^2\right)=\frac{2}{d}\sum_{i=1}^{\frac{d-1}{2}}\left|l_i(0)\right|^{2k}\left|l_i(1)\right|^{2(n-k)}.\nonumber
\end{align}
Next, we give the expression for \textbf{even} $d$. Similarly to the odd $d$ case, all the cross terms sum to $0$, but here we have an additional term, due to $\frac{(-1)^E}{d}l_{\frac{d}{2}}^k(0)l_{\frac{d}{2}}^{n-k}(1)$ in Eq.~(\ref{eq:app_evendProbabilities}). 
\begin{align}
       \sum_{E=0}^{d-1} \left(P(E|k)-\frac{1}{d}\right)^2=\frac{2}{d}\sum_{i=1}^{\frac{d-2}{2}}\left|l_i(0)\right|^{2k}\left|l_i(1)\right|^{2(n-k)}+\frac{1}{d}\left(l_{\frac{d}{2}}(0)\right)^{2k}\left(l_{\frac{d}{2}}(1)\right)^{2(n-k)}.
\end{align}
The $\frac{d}{2}$ term has a different factor from the rest. However, this does not make any difference, since we are only interested in the bounds in the asymptotic limit of $n\rightarrow\infty$, and the summation over $k$ has the same form as for the rest of the terms $i\in\left\{1,2,\dots,\frac{d-2}{2}\right\}$:
\begin{align}
     \sum_{k=0}^{n}\binom{n}{k}\left(l_{\frac{d}{2}}(0)\right)^{2k}\left(l_{\frac{d}{2}}(1)\right)^{2(n-k)}=\left(\left(l_{\frac{d}{2}}(0)\right)^{2}+\left(l_{\frac{d}{2}}(1)\right)^{2}\right)^n.
\end{align}

\section{Proof of Result~\ref{th:ml}}\label{app:ml}
This section of the Appendix includes technical derivation of the proof of Result~\ref{th:ml}. First part is devoted to identifying the set $\mathrm{Q}_1$ from Ref.~\cite{navascues2007bounding} and to precise description of the corresponding moment matrix $\Gamma$. In the second part, we use the symmetries of the derived inequalities to transform  $\Gamma$ to  $\tilde\Gamma$ and then block-diagonalize it.

\subsection{Macroscopic Locality and the corresponding $\Gamma$ matrix}\label{app:subsecion_B1}
Ref.~\cite{navascues2007bounding} defines a hierarchy, often referred to as the NPA hierarchy, of necessary conditions for the observed correlations $\{\Pr(A=k,B=l|\alpha=i,\beta=j)\}_{k,l,i,j}$ to have a quantum-mechanical realization in terms of positive operator-valued measure (POVM) elements of Alice $\{\Al^i_k\}_{i,k}$, and Bob $\{\Bl^j_l\}_{j,l}$ and a bipartite quantum state $\rho_{AB}$. This hierarchy is also sufficient for the systems of finite dimension. For every level of this hierarchy, the condition is expressed in terms of the so-called moment matrix $\Gamma$, in which some of the terms are specified by $\Pr(A=k,B=l|\alpha=i,\beta=j)$, and which is required to be positive semidefinite.
Generally, to define $\Gamma$, one considers a set of operators $\mathcal{O}$, which includes the POVM elements $\{\Al^i_k\}_{i,k}$, $\{\Bl^j_l\}_{j,l}$ and their products. Fixing an order of operators $\mathcal{O}$, one constructs a moment matrix $\Gamma$, such that an element on $i$-th row and $j$-th column of $\Gamma$ is associated to the expectation value of the product of $i$-th and $j$-th operators in $\mathcal{O}$.

In the first level of the hierarchy of Ref.~\cite{navascues2007bounding}, which is equivalent to Macroscopic Locality, applied to our problem, the set of operators $\mathcal{O}$ is simply 
\begin{equation}\label{eq:app_O}
  \mathcal{O}=\{\mathds{1}\}\cup \{ \Al^i_0,\dots,\Al^i_{d-1}\}_{i=0}^{d-1}\cup \{ \Bl^j_0,\dots,\Bl^j_{d-1}\}_{j=0}^{1}, 
\end{equation}
where $\mathds{1}$ is the identity operator defined in the same space as the POVM elements $\{\Al^i_k\}_{i,k}$, $\{\Bl^j_l\}_{j,l}$. It is most convenient to identify the elements of matrix $\Gamma$ by the corresponding operators, i.e., if $\mathcal{O}_i$, $\mathcal{O}_j$ are the $i$-th and $j$-th elements of the set $\mathcal{O}$, then the element of the matrix $\Gamma$ on $i$-th row and $j$-th column we denote as $\Gamma(\mathcal{O}_i,\mathcal{O}_j)$. For the set of operators $\mathcal{O}$ in Eq.~(\ref{eq:app_O}), the square matrix $\Gamma$ is $(d+1)^2 \times (d+1)^2$.

Some of the terms of $\Gamma$ are fixed by $\Pr(A=k,B=l|\alpha=i,\beta=j)$, in particular
\begin{equation}
    \begin{split}
        \Gamma(\mathds{1},\mathds{1}) &= 1,\\
        \Gamma(\mathds{1},\Al^i_k) &= \Gamma(\Al^i_k,\Al^i_k) = \Pr(A=k|\alpha=i),\quad \forall i,k\in[d],\\
        \Gamma(\mathds{1},\Bl^j_l) &= \Gamma(\Bl^j_l,\Bl^j_l) =  \Pr(B=l|\beta=j),\quad \forall l\in[d],j\in[2],\\
        \Gamma(\Al^i_k\,,\Al^{i}_{k'}) &= 0,\quad\forall k'\neq k,\forall i\in [d],\\
        \Gamma(\Bl^j_l\,,\Bl^j_{l'}) &= 0,\quad\forall l'\neq l,\forall j\in [2],\\ 
        \Gamma(\Al^i_k\,,B^j_l) &= \Pr(A=k,B=l|\alpha=i,\beta=j),\quad \forall i,k,l\in[d],j\in[2],
    \end{split}
\end{equation}
while the rest, namely $\Gamma(\Al^i_k\,,\Al^{i'}_{k'})$, for $i'\neq i$, and $\Gamma(\Bl^j_l\,,\Bl^{j'}_{l'})$ for $j'\neq j$, are the variables of optimization. If one can find an assignment to all $\Gamma(\Al^i_k\,,\Al^{i'}_{k'})$ and $\Gamma(\Bl^j_l\,,\Bl^{j'}_{l'})$ such that $\Gamma\geq 0$, then the correlations $\Pr(A=k,B=l|\alpha=i,\beta=j)$ satisfy the Macroscopic Locality principle~\cite{navascues2010glance}.

In the $d2dd$ scenario, which we consider in this paper, the moment matrix $\Gamma$ for the first level of NPA hierarchy can be written as
\begin{equation}\label{app:eq_Gamma}
    \Gamma=
\begin{pmatrix}1 & \bra{P_{A}^{0}} & \dots & \bra{P_{A}^{d-1}} & \bra{P_{B}^{0}} & \bra{P_{B}^{1}}\\
\ket{P_{A}^{0}}  & V^{0,0} & \dots & V^{0,d-1} & P^{0,0} & P^{0,1}\\
\vdots & \vdots & \ddots & \vdots & \vdots & \vdots\\
\ket{P_{A}^{d-1}}  & (V^{0,d-1})^T & \dots & V^{d-1,d-1} & P^{d-1,0} & P^{d-1,1}\\
\ket{P_{B}^{0}}  & (P^{0,0})^{T} & \dots & (P^{d-1,0})^{T} & W^{0,0} & W^{0,1}\\
\ket{P_{B}^{1}}  & (P^{0,1})^{T} & \dots & (P^{d-1,1})^{T} & (W^{0,1})^T &  W^{1,1}
\end{pmatrix},
\end{equation}
where vectors $\ket{P^i_A}$, $\ket{P^j_B} \in \mathds{R}^d$ are defined as
\begin{equation}
\begin{split}
   \ket{P^i_A} &= \sum_{k=0}^{d-1}\Pr(A=k|\alpha=i)\,\ket{k},\quad \forall i\in[d],\qquad  \ket{P^j_B} = \sum_{l=0}^{d-1}\Pr(B=l|\beta=j)\,\ket{l},\quad \forall j\in[2].
 \end{split}
\end{equation}
The matrices $V^{i,i'},P^{i,j}$, and $W^{j,j'}$, which are all $d\times d$, are defined below
\begin{equation}
\begin{split}
   P^{i,i'} &= \sum_{k,l}^{d-1}\Pr(A=k,B=l|\alpha=i,\beta=j)\kb{k}{l},\quad \forall i\in[d],j\in[2],\\
   V^{i,i'} &= \sum_{k,k'=0}^{d-1}v^{i,i'}_{k,k'}\kb{k}{k'},\quad\forall i< i',\qquad V^{i,i} = \sum_{k=0}^{d-1}\Pr(A=k|\alpha=i)\kb{k}{k},\quad \forall i\in[d],\\
      W^{0,1} &= \sum_{l,l'}^{d-1}w_{l,l'}\kb{l}{l'},\quad W^{j,j} = \sum_{l=0}^{d-1}\Pr(B=l|\beta=j)\kb{l}{l},\quad \forall j\in[2].
 \end{split}
\end{equation}
In the above formulas, we denoted the variables of optimization as $v^{i,i'}_{k,k'}$ and $w_{l,l'}$, which correspond to the following elements of matrix $\Gamma$
\begin{equation}
    v^{i,i'}_{k,k'} = \Gamma(\Al^{i}_{k}\,,\Al^{i'}_{k'}),\quad w_{l,l'} = \Gamma(\Bl^0_l\,,\Bl^1_{l'}),\quad \forall i,i',k,k',l,l'\in[d],\; i < i'.
\end{equation}
Due to the relations $\sum_{k=0}^{d-1}\Al^i_k = \openone$, $\forall i\in[d]$, $\sum_{l=0}^{d-1}\Bl^j_l = \openone$, $\forall j\in[2]$ on POVM elements there are further constraints on $v^{i,i'}_{k,k'}$ and $w_{l,l'}$, namely
\begin{equation}
\begin{split}
    \sum_{k=0}^{d-1}v^{i,i'}_{k,k'} &= \Pr(A=k'|\alpha=i'),\quad \sum_{k'=0}^{d-1}v^{i,i'}_{k,k'} = \Pr(A=k|\alpha=i),\quad \forall i,i',k,k'\in[d],\; i < i'. \\
       \sum_{l=0}^{d-1}w_{l,l'} &= \Pr(B=l'|\beta=1),\quad \sum_{l'=0}^{d-1}w_{l,l'} = \Pr(B=l|\beta=0),\quad \forall l,l'\in[d].
    \end{split}
\end{equation}

We are interested in characterizing the set of ML correlations in the space of probabilities $\{p(e|j)\}_{e,j}$ from Eq.~(\ref{eq:pej}). Here, for the convenience we restate the form of these probabilities,
\begin{equation}\label{eq:app_pej}
        p(e|j)=\frac{1}{d}\sum_{i=0}^{d-1}\Pr(A \oplus B = i\cdot j\oplus e|\alpha=i,\beta=j),\quad \forall e\in[d],j\in[2].
\end{equation}
Since the probabilities $\{p(e|j)\}_{e,j}$ form a subspace of the entire space of correlations $\Pr(A=k,B=l|\alpha=i,\beta=j)$, we can reduce the size of the problem by considering the symmetries present in $\{p(e|j)\}_{e,j}$.

\subsection{Symmetrization and block-diagonalization of $\Gamma$}\label{app:subsecion_B2}
In our proof, we consider the cyclic group of permutations and not the whole symmetric group $S_d$. Let $\pi_A$ denote the generating element of cyclic permutation of outcomes of Alice, and similarly $\pi_B$ for permutation of outcomes of Bob, and $\pi_\alpha$ for permutation of settings of Alice. We define the action of these permutations on POVM elements of Alice and Bob as follows,
\begin{equation}\label{eq:app_g_action}
    \pi_A^r . \Al^i_k = \Al^i_{k\oplus r},\quad \pi^r_B . \Bl^j_l = \Bl^j_{l\oplus r},\quad \pi^r_\alpha . \Al^i_k = \Al^{i\oplus r}_k, \quad \forall r,i,k,l\in [d],j\in[2],
\end{equation}
where $\pi_A^r$ is the $r$-th power of the permutation $\pi_A$, and the same notation is used for $\pi_B$ and $\pi_\alpha$. The order of each of these elements is $d$, hence the inverse of e.g., $\pi_A^r$ is $\pi_A^{(d-r)}$. Finally, let $\pi_{AB} = \pi_A\cdot\pi_B^{-1}$ be the composition of permutation of outcomes of Alice with the inverse permutation of outcomes of Bob, and $\pi_{\alpha B}=\pi_\alpha\cdot\pi^j_B$ be the composition of permutation of Alice's settings combined with permutation of Bob's outputs, conditioned on $j$. We can now define the commutative symmetry group 
\begin{equation}
    G = \langle \{\pi_{AB},\pi_{\alpha B}\}| \pi_{AB}\cdot\pi_{\alpha B} = \pi_{\alpha B}\cdot\pi_{AB}\rangle.
\end{equation}

The action of the symmetry group $G$ can be extended to elements of the moment matrix $\Gamma$, since we enumerate its rows and columns by the POVM elements. More precisely, we write
\begin{equation}\label{eq:app_g_act_g}
\begin{split}
    (\pi_{AB}^r\cdot\pi^s_{\alpha B}) . \Gamma(\mathds{1},\mathds{1}) & = \Gamma(\mathds{1},\mathds{1}),\quad \forall r,s\in[d],\\
    (\pi_{AB}^r\cdot\pi^s_{\alpha B}) . \Gamma(\mathds{1},\Al^i_k) & = \Gamma(\mathds{1},\Al^{i\oplus s}_{k\oplus r}),\quad \forall r,s,i,k\in[d],\\
    (\pi_{AB}^r\cdot\pi^s_{\alpha B}) . \Gamma(\mathds{1},\Bl^j_l) & = \Gamma(\mathds{1},\Bl^{j}_{l\ominus r\oplus s\cdot j}),\quad \forall r,s,l\in[d], j\in[2],\\
        (\pi_{AB}^r\cdot\pi^s_{\alpha B}) . \Gamma(\Al^i_k\,,\Al^{i'}_{k'}) & = \Gamma(\Al^{i\oplus s}_{k\oplus r}\,,\Al^{i'\oplus s}_{k'\oplus r}),\quad\forall r,s,i,i',k,k'\in[d],\\
        (\pi_{AB}^r\cdot\pi^s_{\alpha B}) . \Gamma(\Bl^j_l\,,\Bl^{j'}_{l'}) & = \Gamma(\Bl^{j}_{l\ominus r\oplus s\cdot j}\,,\Bl^{j'}_{l'\ominus r\oplus s\cdot j'}),\quad \forall r,s,l,l'\in[d],j,j'\in[2],\\
        (\pi_{AB}^r\cdot\pi^s_{\alpha B}) . \Gamma(\Al^i_k\,,\Bl^j_l) & = \Gamma(\Al^{i\oplus s}_{k\oplus r}\,,\Bl^{j}_{l\ominus r\oplus s\cdot j}),\quad \forall r,s,i,k,l\in[d],j\in[2].
    \end{split}
\end{equation}

The notation $(\pi_{AB}^r\cdot\pi^s_{\alpha B}).\Gamma$ should be understood as the action of $(\pi_{AB}^r\cdot\pi^s_{\alpha B})$ on each element of the matrix $\Gamma$, as defined by Eq.~(\ref{eq:app_g_act_g}). 
Since we already identified probabilities $\Pr(A=i,B=j|\alpha=k,\beta=l)$ with elements $\Gamma(\Al^k_i,\Bl^l_j)$ of matrix $\Gamma$, the action of $G$ extends to the former in the obvious way:
\begin{equation}
   (\pi_{AB}^r\cdot\pi^s_{\alpha B}).\Pr(A=k,B=l|\alpha=i,\beta=j) = \Pr(A=k\oplus r,B=l\ominus r\oplus s\cdot j|\alpha=i\oplus s,\beta=j), 
\end{equation}
$\forall r,s,i,k,l\in [d],j\in[2].$ Form the above Eq. it must be clear that $G$ preserves the probabilities $p(e|j)$ in Eq.~(\ref{eq:app_pej}).

 In our proof, we define the following matrix
 \begin{equation}\label{eq:app_av_g}
    \tilde{\Gamma}=\frac{1}{d}\sum_{r,s=0}^{d-1} (\pi_{AB}^r\cdot\pi^s_\alpha) . \Gamma,
 \end{equation}
 and as follows from Ref.~\cite{rosset2015characterization}, the condition of $\tilde{\Gamma}\geq 0$ is equivalent to the condition $\Gamma\geq 0$ if one is concerned with bounding the space of probabilities $\{p(e|j)\}_{e,j}$ in Eq.~(\ref{eq:app_pej}).
 
In Eq.~(\ref{eq:app_g_act_g}) we determined what happens to the entries of $\Gamma$ matrix when we apply the permutation from $G$. These relations can be used to determine the orbits corresponding to each element of the moment matrix, and subsequently to find the orbit mean values, when we sum over $r$ and $s$ in Eq.~(\ref{eq:app_av_g}).
\begin{equation}\label{eq:app_g_sum_g}
\begin{split}
    \sum_{r,s=0}^{d-1}\Gamma(\mathds{1},\mathds{1}) &= d^2, \quad         \sum_{r,s=0}^{d-1}\Gamma(\Bl^{0}_{l\ominus r}\,,\Bl^{1}_{l'\ominus r\oplus s}) = 1,\quad \forall l,l'\in[d],\\
    \sum_{r,s=0}^{d-1}\Gamma(\Al^{i\oplus s}_{k\oplus r}\,,\Al^{i'\oplus s}_{k'\oplus r}) & = \sum_{r,s=0}^{d-1}v^{s,i'\ominus i\oplus s}_{r,k'\ominus k\oplus r},\quad\forall i<i',\; i,k,k'\in[d],\\
    \sum_{r,s=0}^{d-1}\Gamma(\mathds{1},\Al^{i\oplus s}_{k\oplus r}) &= d,\quad  \sum_{r,s=0}^{d-1}\Gamma(\Al^{i\oplus s}_{k\oplus r}\,,\Al^{i\oplus s}_{k\oplus r})= d,\quad \forall i,k\in[d],\\
    \sum_{r,s=0}^{d-1}\Gamma(\mathds{1},\Bl^{j}_{l\ominus r\oplus s\cdot j}) & = d,\quad  \sum_{r,s=0}^{d-1}\Gamma(\Bl^{j}_{l\ominus r\oplus s\cdot j}\,,\Bl^{j}_{l\ominus r\oplus s\cdot j}) = d,\quad \forall l\in[d],j\in[2],
    \end{split}
\end{equation}
and most importantly,
\begin{equation}
            \sum_{r,s=0}^{d-1}\Gamma(\Al^{i\oplus s}_{k\oplus r}\,,\Bl^{j}_{l\ominus r\oplus s\cdot j}) =\sum_{r,s=0}^{d-1}\Pr(A=r,B=l\oplus k\ominus r\oplus j\cdot(s\ominus i)|\alpha=s,\beta=j)=d p(l\oplus k\ominus i\cdot j|j),
\end{equation}
where the last Eq. is valid for all $i,k,l\in[d],j\in[2]$.

 Similar to the way we defined $\Gamma$ in Eq.~(\ref{app:eq_Gamma}), we can define $\tilde{\Gamma}$ in terms of $d\times d$ blocks. Using the notation $\ket{+}=\sum_{i=0}^{d-1}\ket{i}$, we can express the new matrix as follows
\begin{equation}\label{app:eq_tilde_gamma}
\tilde{\Gamma}=\begin{pmatrix}d & \bra{+} & \bra{+} & \dots & \bra{+} & \bra{+} & \bra{+}\\
\ket{+} & \openone & \tilde{V}^{1} & \dots & \tilde{V}^{d-1} & \tilde{P}^{0,0} & \tilde{P}^{0,1}\\
\ket{+} & (\tilde{V}^{1})^{\text{T}} & \openone & \dots & \tilde{V}^{d-2} & \tilde{P}^{1,0} & \tilde{P}^{1,1}\\
\vdots & \vdots & \vdots & \ddots & \vdots & \vdots & \vdots\\
\ket{+} & (\tilde{V}^{d-1})^{T} & (\tilde{V}^{d-2})^{T} & \dots & \openone & \tilde{P}^{d-1,0} & \tilde{P}^{d-1,1}\\
\ket{+} & (\tilde{P}^{0,0})^{T} & (\tilde{P}^{1,0})^{T} & \dots & (\tilde{P}^{d-1,0})^{T} & \openone & \mathbbm{J}/d\\
\ket{+} & (\tilde{P}^{0,1})^{T} & (\tilde{P}^{1,1})^{T} & \dots & (\tilde{P}^{d-1,1})^{T} & \mathbbm{J}/d & \openone
\end{pmatrix}.
\end{equation}
 The sub-matrices $\tilde{V}^{\Delta i}$, where $1\leq \Delta i \leq d-1$ are defined as follows,  
\begin{equation}
    \tilde V^{\Delta i}=\sum_{\Delta k} \tilde v^{\Delta i}_{\Delta k} {X}^{\Delta k}, \quad \mbox{ where } \quad \tilde{v}^{\Delta i}_{\Delta k}= \frac{1}{d}\sum_{r,s=0}^{d-1}v^{s,\Delta i\oplus s}_{r,\Delta k\oplus r}.
\end{equation}

Finally, we consider $\tilde{P}^{i,j}$. 
\begin{equation}
    \tilde{P}^{i,j}= \sum_{e=0}^{d-1}S_{(e+i \cdot j)} p(e|j),
\end{equation}
where we remind the reader that $S_E = \sum_{k=0}^{d-1}\kb{k}{\bar{k}\oplus E}$.
In order to further simplify $\tilde{\Gamma}$, we apply a unitary transformation $U$, specified below.
\begin{align} \label{app:eq_unitary}
  \Gamma'=  U^\dagger\tilde\Gamma U&=\begin{pmatrix}1\\
 & H^\dagger\\
 &  & \ddots\\
 &  &  & H^\dagger\\
 &  &  &  &H^\dagger S^\dagger_{d-1}\\
 &  &  &  &  & H^\dagger S^\dagger_{d-1}
\end{pmatrix}\cdot \tilde \Gamma\cdot
\begin{pmatrix}1\\
 & H\\
 &  & \ddots\\
 &  &  & H\\
 &  &  &  & S_{d-1}H\\
 &  &  &  &  & S_{d-1}H
\end{pmatrix}\\[11pt]
=& \begin{pmatrix}d & \sqrt{d}\bra{0} & \dots & \sqrt{d}\bra{0} & \sqrt{d}\bra{0} & \sqrt{d}\bra{0}\\
\sqrt{d}\ket{0} & \openone & \dots & H^{\dagger}\tilde{V}^{( d-1)}H & H^{\dagger}\tilde{P}^{0,0}S_{d-1}H & H^{\dagger}\tilde{P}^{0,1}S_{d-1}H\\
\vdots & \vdots & \ddots & \vdots & \vdots & \vdots\\
\sqrt{d}\ket{0} & H^{\dagger}(\tilde{V}^{(d-1)})^{T}H & \dots & \openone & H^{\dagger}\tilde{P}^{d-1,0}S_{d-1}H & H^{\dagger}\tilde{P}^{d-1,1}S_{d-1}H\\
\sqrt{d}\ket{0} & H^{\dagger}S_{d-1}^{\dagger}(\tilde{P}^{0,0})^{T}H & \dots & (\tilde{P}^{d-1,0})^{T}H & \openone & \frac{1}{d}H^{\dagger}S_{d-1}^{\dagger}\mathbbm{J}S_{d-1}H\\
\sqrt{d}\ket{0} & H^{\dagger}S_{d-1}^{\dagger}(\tilde{P}^{0,1})^{T}H & \dots & (\tilde{P}^{d-1,1})^{T}H & \frac{1}{d}H^{\dagger}S_{d-1}^{\dagger}\mathbbm{J}S_{d-1}H & \openone
\end{pmatrix}.
\end{align}
We can separately analyze the resulting blocks and see that the unitary action $U$ diagonalizes all of them, $\frac{1}{d}  H^{\dagger}S^{\dagger}_{d-1}\mathbbm{J} S_{d-1}H = \frac{1}{d} H^{\dagger}\mathbbm{J} H=\kb{0}{0}$ and
\begin{align}
  H^{\dagger}\tilde{V}^{\Delta i}H =  &  H^{\dagger}\left(\sum_{\Delta k} \tilde v^{\Delta i}_{\Delta k} {X}^{\Delta k}\right)H=\sum_{\Delta k} \tilde v^{\Delta i}_{\Delta k} H^{\dagger}X^{\Delta k}H=\sum_{\Delta k} \tilde v^{\Delta i}_{\Delta k} Z^{\Delta k}\equiv \sum_{k=0}^{d-1}\nu_m^{ i}\kb{m}{m}.\label{app:eq_vcoeff}\\
 H^\dagger \tilde{P}^{i,j}S_{d-1}H =   &  H^\dagger \left(\sum_{e=0}^{d-1}S_{(e+i \cdot j)} p(e|j)\right)S_{d-1}H =\sum_{e=0}^{d-1}p(e|j) H^\dagger \left(S_{(e+i \cdot j)} S_{d-1}\right)H\\
 &=\sum_{e=0}^{d-1}p(e|j) H^\dagger X^{(e+i \cdot j+1)}H=\sum_{e=0}^{d-1}p(e|j)Z^{(e+i \cdot j+1)}\equiv \sum_{k=0}^{d-1}\omega^{m\cdot (i\cdot j)}p_m^{j}\kb{m}{m}, \label{app:eq_pcoeff}
\end{align}
where $p^j_m = \sum_{e=0}^{d-1}\omega^{m\cdot(1+e)}p(e|j)$. Moreover, from the normalization of optimization variables and of probabilities we have that $\nu_0^i=1$ and $p_0^{j}=1$ for all $i\in [d],j\in [2]$.  Having diagonalized all the matrices, we can rearrange $\Gamma'$ in a block-diagonal form. The first block can be constructed by concatenating nonzero entries of first row and first column and by picking only the first $(1,1)$ entry of each $d\times d$ sub-block of $\Gamma'$. The resulting block can be expressed as follows,
\begin{equation}\label{app:eq_gamma_0}
 \Gamma_0=   
\begin{pmatrix}d & \sqrt{d} & \sqrt{d} & \dots & \sqrt{d} & \sqrt{d} & \sqrt{d}\\
\sqrt{d} & 1 & \nu_{0}^{1} & \dots & \nu_{0}^{d-1} & p_{0}^{0} & p_{0}^{1}\\
\sqrt{d} & (\nu_{0}^{1})^{\dagger} & 1 & \ddots & \vdots & p_{0}^{0} & p_{0}^{1}\\
\vdots & \vdots & \ddots & \ddots & \nu_{0}^{1} & \vdots & \vdots\\
\sqrt{d} & (\nu_{0}^{d-1})^{\dagger} & \dots & (\nu_{0}^{1})^{\dagger} & 1 & p_{0}^{0} & p_{0}^{1}\\
\sqrt{d} & (p_{0}^{0})^{\dagger} & \dots &  & (p_{0}^{0})^{\dagger} & 1 & 1\\
\sqrt{d} & (p_{0}^{1})^{\dagger} & \dots &  & (p_{0}^{1})^{\dagger} & 1 & 1
\end{pmatrix}=\begin{pmatrix}d & \sqrt{d} & \sqrt{d} & \dots & \sqrt{d} & \sqrt{d} & \sqrt{d}\\
\sqrt{d} & 1 & 1 & \dots & 1 & 1 & 1\\
\sqrt{d} & 1 & 1 & \ddots & \vdots & 1 & 1\\
\vdots & \vdots & \ddots & \ddots & 1 & \vdots & \vdots\\
\sqrt{d} & 1 & \dots & 1 & 1 & 1 & 1\\
\sqrt{d} & 1 & \dots & 1 & 1 & 1 & 1\\
\sqrt{d} & 1 & \dots & 1 & 1 & 1 & 1
\end{pmatrix}.
\end{equation}

Matrix $\Gamma_0\geq 0$ (is positive semidefinite). The next block, $\Gamma_1$ can be constructed by picking only the second diagonal entry $(2,2)$ from each sub-block of $\Gamma'$.
And more generally, $(m-1)$-th block can be constructed by picking only the diagonal entry $(m,m)$ from each sub-block of $\Gamma'$ for $1\leq m\leq d+1$:

\begin{equation}\label{app:eq_blockMatrix}
    \Gamma_m=\begin{pmatrix}1 & \nu_{m}^{1} & \dots & \nu_{m}^{d-1} & p_{m}^{0} & p_{m}^{1}\\
(\nu_{m}^{1})^{\dagger} & 1 & \ddots & \vdots & p_{m}^{0} & \omega ^{m}p_{m}^{1}\\
\vdots & \ddots & \ddots & \nu_{m}^{1} & \vdots & \vdots\\
(\nu_{m}^{d-1})^{\dagger} & \dots & (\nu_{m}^{1})^{\dagger} & 1 & p_{m}^{0} &  \omega ^{(d-1)\cdot m}p_{m}^{1}\\
(p_{m}^{0})^{\dagger} & \dots &  & (p_{m}^{0})^{\dagger} & 1 & 0\\
(p_{m}^{1})^{\dagger} & \dots &  & ( \omega ^{(d-1)\cdot m}p_{m}^{1})^{\dagger} & 0 & 1
\end{pmatrix}.
\end{equation}

We have, thus, reduced checking compatibility with ML to  positivity of block-diagonal matrices.
\end{appendix}

\bibliographystyle{unsrtnat}
\bibliography{bib}

\end{document}